\documentclass[11pt,a4paper]{article}
\usepackage{longtable}
\usepackage{amsmath}
\usepackage{graphicx}
\usepackage{bm}
\usepackage{footmisc}
\usepackage{afterpage}

\newcommand*{\no}{\noindent}
\newcommand*{\bea}{\begin{eqnarray}}
\newcommand*{\eea}{\end{eqnarray}}
\newcommand*{\be}{\begin{equation}}
\newcommand*{\ee}{\end{equation}}

\newcommand*{\pref}[1]{(\ref{#1})}

\newcommand*{\mn}{{\mu\nu}}

\newcommand*{\nn}{\nonumber}

\newcommand*{\tr}{\mathrm{tr}}

\usepackage[square,comma,sort&compress,numbers]{natbib}

\title{Propagators and topology}

\author{Axel Maas\\
Institute of Physics, University of Graz,\\
Universit\"atsplatz 5, A-8010 Graz, Austria}

\begin{document}

\maketitle

\begin{abstract}
Two popular perspectives on the non-perturbative domain of Yang-Mills theories are either in terms of the gluons themselves or in terms of collective gluonic excitations, i.\ e.\ topological excitations. If both views are correct, then they are only two different representations of the same underlying physics. One possibility to investigate this connection is by the determination of gluon correlation functions in topological background fields, as created by the smearing of lattice configurations. This is performed here for the minimal Landau gauge gluon propagator, ghost propagator, and running coupling, both in momentum and position space for SU(2) Yang-Mills theory. The results show that the salient low-momentum features of the propagators are qualitatively retained under smearing at sufficiently small momenta, in agreement with an equivalence of both perspectives. However, the mid-momentum behavior is significantly affected. These results are also relevant for the construction of truncations in functional methods, as they provide hints on necessary properties to be retained in truncations.
\end{abstract}

\section{Introduction}

The non-perturbative, low-energy domain of Yang-Mills theory, and in extension of QCD, remains an interesting conceptual challenge, even if approaches like, e.\ g., lattice gauge theory, functional methods, or chiral perturbation theory already permit to determine many quantities of practical interest. One of the central questions has been for a very long time what the effective degrees of freedom at low energies are. Two perspectives are in terms of the gauge bosons themselves \cite{Maas:2011se,Alkofer:2000wg,Fischer:2006ub,Binosi:2009qm,Boucaud:2011ug,Vandersickel:2012tg} and in terms of collective, i.\ e.\ topological, excitations \cite{Greensite:2003bk,Ripka:2003vv,Negele:2004hs,DiGiacomo:2008nt,Alexandru:2005bn,Bornyakov:2013iva}, like (center) vortices \cite{Greensite:2003bk}, monopoles \cite{Greensite:2003bk,Ripka:2003vv,DiGiacomo:2008nt}, instantons \cite{Schafer:1996wv,Ilgenfritz:2012aa}, calorons \cite{Ilgenfritz:2006ju,Kraan:1998pm}, merons \cite{Zimmermann:2012zi}, and dyons \cite{Diakonov:2010qg,Bornyakov:2008im}. These topological configurations are likely not all independent, but intricately related \cite{Greensite:2003bk,deForcrand:2000pg,Boyko:2006ic,Reinhardt:2001kf,Caudy:2007sf}.

Provided both views are correct, they are necessarily just two different representations of the same physics. Since for both views a plethora of evidence exists, this appears likely to be the case. In fact, for simpler models such relations are explicitly known \cite{Alkofer:2000wg,Radozycki:1998xs}. Thus, it should be possible to establish this relation explicitly, a challenge which remains so far unsolved \cite{Alkofer:2006fu}. However, since both, gluons and gluonic excitations, are inherently gauge-dependent, any such relation could be itself gauge-dependent. Still, it would be significant progress to establish the details of this connection at least for one gauge. There have been a number of investigations contributing to this endeavor, in both Coulomb and Landau gauge \cite{Greensite:2004ur,Gattnar:2004bf,Maas:2011se,Maas:2005qt,Maas:2006ss,Langfeld:2002dd,Boucaud:2003xi,Maas:2008uz,Langfeld:2001cz,Quandt:2010yq,Chernodub:2011pr}\footnote{There are also investigations on the quark propagator \cite{Bowman:2010zr,Bowman:2008qd} and indirect investigations of topology-sensitive hadronic observables \cite{Alkofer:2008et}, but this is outside the focus of this work.}, with respect to different types of topological excitations. These investigations provided evidence that such a link indeed exists, and that the most characteristic low-momentum features of gluonic correlation functions are likely reflecting (or formed by, depending on perspective) topological excitations.

Lattice calculations, using smearing \cite{Gattringer:2010zz,Bruckmann:2006wf,DeGrand:1997ss}, provide a tool to isolate from field configurations the (self-dual) topological part. This permits another way to determine correlation functions of gluons in a topological field configuration, i.\ e.\ how the gluons inside a collective gluonic excitation behave. This possibility has so far only been explored in preliminary investigations \cite{Maas:2008uz,Maas:2011se}. Here, this will be extended to a full systematic investigation of this possibility in (minimal) Landau gauge for the gluon, the corresponding ghost, and the running coupling, both in momentum and position space. Because these investigations are computationally much more expensive than calculating just the propagators, this will be only possible over a limited range of lattice settings, and therefore will not yet answer questions about the deep infrared. However, they are complementary to investigations using center projections \cite{Gattnar:2004bf,Langfeld:2002dd,Langfeld:2001cz,Quandt:2010yq}, and therefore offer a novel perspective on the interplay of collective and single gluon excitations at low energies.

The technical details of these investigations are briefly described in section \ref{stech}. A discussion of the selection criteria for the background configurations on which to measure the propagators is given in section \ref{ssel}. Results for the gluon are then presented separately for momentum space and position space in sections \ref{smom} and \ref{spos}, respectively, in the results section \ref{sres}. The ghost propagator is investigated in section \ref{sghost}. Section \ref{sres} also contains results on further derived quantities. The results are summarized in section \ref{sconc}, after a speculative interpretation in section \ref{sspec}. There, are also a few remarks will be made on how these results can be exploited to ensure truncations of functional equations \cite{Maas:2011se,Alkofer:2000wg,Binosi:2009qm,Boucaud:2011ug,Vandersickel:2012tg,Fischer:2006ub} adequately capture the contributions from collective excitations.

\section{Technical details}\label{stech}

\begin{table}
\caption{\label{configs}The configurations employed. $N$ is the (symmetric) extent of the lattice of total volume $N^4$. The lattice spacing at given gauge coupling $\beta$ has been determined using the data of \cite{Fingberg:1992ju}, setting the string tension to $(440$ MeV$)^2$. The number of configurations before and after the slash are the number used for the short and long cooling, respectively, see text. Therm.\ and sweeps are the number of configurations dropped for thermalization and between two measurements. To reduce correlations, prevalent for topological quantities \cite{DelDebbio:2002xa}, typical ${\cal O}(100)$ independent runs have been performed for each lattice setting. Long gives the number of total APE sweeps performed for the self-dual configurations, while Int.\ gives the measurement interval between APE sweeps in this case.}
\begin{tabular}{|c|c|c|c|c|c|c|c|c|}
\hline
$\beta$ & $N$ & $a^{-1}$ [GeV] & $L=aN$ [fm] & Configurations & Therm.\ & Sweeps & Long & Int.\ \cr
\hline
2.2 & 8 & 0.94 & 1.7 & 11513/11554 & 380 & 38 & 400 & 10 \cr
\hline
2.2 & 12 & 0.94 & 2.5 & 12716/12179 & 420 & 42 & 440 & 11 \cr
\hline
2.2 & 16 & 0.94 & 3.4 & 2304/3798 & 460 & 46 & 484 & 12 \cr
\hline
2.2 & 20 & 0.94 & 4.2 & 2718/2494 & 500 & 50 & 532 & 13 \cr
\hline
2.2 & 24 & 0.94 & 5.0 & 2852/2200 & 540 & 54 & 585 & 14 \cr
\hline
2.35 & 8 & 1.4 & 1.1 & 11196/1555 & 380 & 38 & 400 & 10 \cr
\hline
2.35 & 12 & 1.4 & 1.7 & 12079/12493 & 420 & 42 & 440 & 11 \cr
\hline
2.35 & 16 & 1.4 & 2.2 & 2585/3016 & 460 & 46 & 484 & 12 \cr
\hline
2.35 & 20 & 1.4 & 2.8 & 3722/2679 & 500 & 50 & 532 & 13 \cr
\hline
2.35 & 24 & 1.4 & 3.4 & 2662/1989 & 540 & 54 & 585 & 14 \cr
\hline
2.5 & 8 & 2.3 & 0.69 & 11501/10970 & 380 & 38 & 400 & 10 \cr
\hline
2.5 & 12 & 2.3 & 1.0 & 12220/6786 & 420 & 42 & 440 & 11 \cr
\hline
2.5 & 16 & 2.3 & 1.4 & 8073/3343 & 460 & 46 & 484 & 12 \cr
\hline
2.5 & 20 & 2.3 & 1.7 & 3869/2654 & 500 & 50 & 532 & 13 \cr
\hline
2.5 & 24 & 2.3 & 2.0 & 2823/1735 & 540 & 54 & 585 & 14 \cr
\hline
\end{tabular}
\end{table}

The lattice configurations have been created using the standard SU(2) Yang-Mills Wilson action \cite{Gattringer:2010zz}, using the hybrid-overrelaxation algorithm described in \cite{Cucchieri:2006tf}. The list of configurations and lattice parameters are given in table \ref{configs}.

There are a number of possibilities to isolate the topological content of the generated lattice configurations. All of these algorithms appear to provide qualitatively similar results, but differ quantitatively by  10-30\% \cite{Bruckmann:2006wf,Bonati:2014tqa}. Since for this first investigation the qualitative effects are most interesting, these deviations are not too important, and hence only a single method will be used. This will be APE-smearing \cite{DeGrand:1997ss} with a lexicographical update. This procedure is equivalent to the mathematically better defined Wilson flow, provided the smearing levels are suitably selected \cite{Bonati:2014tqa}, a subject to be discussed in more detail in section \ref{ssel}.

In the APE smearing process, a single smearing sweep replaces all link variables by the following prescription
\bea
U_\mu(x)&\to& \alpha U_\mu(x)+\frac{1-\alpha}{2(d-1)}\sum_{\nu\neq\mu}\left(U_\nu(x+e_\mu)U^+_\mu(x+e_\nu)U^+_\nu(x)\right.\nn\\
&&\left.\left.+U_\nu^+(x+e_\mu-e_\nu)U_\mu^+(x-e_\nu)U_\nu(x-e_\nu)\right)\right|_\mathrm{projected\;to\;the\;group}\nn,
\eea
\no where ``projected to the group'' implies that the non-group element $U_\mu'$ found after addition is replaced by the group element $U_\mu''$ closest to the result, where the distance is given by $\tr U_\mu'' U_\mu'$, with no summation implied. For the present SU(2) case this can be achieved by a multiplicative factor. The parameter $\alpha$ can be used to tune the smearing. It will be set throughout to 0.55, which is a value being very convenient in many investigations \cite{Bruckmann:2006wf}. Typically, only a few APE smearing sweeps are used in spectroscopy applications. But to smear to an essentially topological content, i.\ e.\ one satisfying approximately the (anti-)self-duality equations $F_\mu=\pm\epsilon_{\mu\nu\rho\sigma} F_{\rho\sigma}$, typical for topological (instanton-like) configurations \cite{Felsager:1981iy}, several more APE smearing sweeps are necessary. Depending on the lattice parameters, this can be a few tens, or several hundred. It should be noted that the smearing is done here in all directions, not just in spatial directions, owing to the aim of creating self-dual configurations with full O(4) symmetry.

There will be two possibilities considered throughout. The first will be short smearing, with 25 or less smearing sweeps. This will eliminate the ultraviolet fluctuations, but should leave most of the infrared structure intact. The second will be a long smearing, where the configurations are moved far into the self-dual regime. Since this depends on the volume, the number of smearing steps is chosen volume-dependent. For the short smearing a measurement will be performed after each smearing sweep, while for the long smearing only in some interval, again depending on the volume, measurements will be performed. These intervals are given in table \ref{configs}.

The measurements of the propagators requires to fix the gauge, which is chosen to be the minimal Landau gauge \cite{Maas:2011se}, using the algorithm described in \cite{Cucchieri:2006tf}. Since the gauge is not preserved under smearing, it was necessary to fix the gauge for each measurement anew. This made this investigation comparatively expensive, and therefore only the limited set of lattice setups, and especially the rather small volumes, listed in table \ref{configs} could be investigated. This also implies that the lowest reachable momenta are still large, even on the largest and coarsest lattice they are ${\cal O}(250)$ MeV. Thus, the far infrared domain, being an actively researched area \cite{Maas:2011se}, could not be investigated. However, the domain most relevant to (hadronic) bound states \cite{Alkofer:2000wg,Fischer:2006ub} is accessible.

The quantities measured are the color-diagonal part\footnote{The color-off-diagonal part is zero in Landau gauge \cite{Maas:2011se}.} of the gluon propagator $D$ and the ghost propagator $D_G$ in momentum space and the gluon propagator also in position space, the latter also denoted as the Schwinger function $\Delta$. The measurement is performed with the methods described in \cite{Cucchieri:2006tf,Maas:2011se}. Renormalization will be performed for the unsmeared propagators at $\mu=2$ GeV with $\mu^2D(\mu^2)=\mu^2D_G(\mu^2)=1$, though this is essentially only needed in figure \ref{fig:d0}. Hence, results shown are unrenormalized, except when stated otherwise.

The propagators suffice to calculate \cite{vonSmekal:1997vx} the running coupling in the miniMOM scheme \cite{vonSmekal:2009ae} as
\be
\alpha(p^2)=\alpha(\mu^2) p^6D(p^2)D_G(p^2)^2\nn.
\ee 
\no Since this is a renormalization-group-invariant quantity, it can be calculated using the unrenormalized propagators, after fixing $\alpha(\mu^2)$ \cite{vonSmekal:1997vx}.

It is finally interesting to see whether the propagators depend on the topological charge. To measure the topological charge, the simplest lattice realization of the continuum topological charge density operator
\be
q(x)=\frac{1}{32\pi^2}\tr\epsilon_{\mu\nu\rho\sigma}F_\mn(x) F_{\rho\sigma}(x),\label{q}
\ee
\no will be used. This will be performed by calculating first the field strength tensor $F_\mn$ at site $x$ from the link variables, and then calculating the product \pref{q}. The full topological charge $Q$ is then obtained by summation
\be
Q=\sum_x q(x)\label{topcharge}.
\ee
\no Since $Q$ is on a finite lattice usually not exactly an integer, the result of the measurement will be projected to the nearest integer. It should be noted that $Q$ does not count the number of, e.\ g., instanton-like objects, but is the net number. Thus, in a configuration with $n_+$ objects of topological charge 1 and $n_-$ objects of topological charge $-1$, $Q=n_+-n_-$. It therefore cannot give insight on the local structure of a configuration, but merely characterizes the vacuum sector.

The topological charge is also used to monitor the smearing process: In the self-dual regime, the quantity \pref{topcharge} forms (almost) integer plateaus as a function of smearing sweeps, and changes only by (almost) integer jumps \cite{DeGrand:1997ss}. The appearance of these plateaus were used to assure that almost all measurements in the long smearing runs were in the self-dual domain.

The distribution in $Q$ is Gaussian centered around $Q=0$. Thus to sample even modestly large $Q$ requires an exponentially large amount of statistics, which restricts this type of investigation only to a very limited set of lattice parameters and/or $Q$ values.

\section{Selection of smearing levels}\label{ssel}

\begin{figure}
\includegraphics[width=\linewidth]{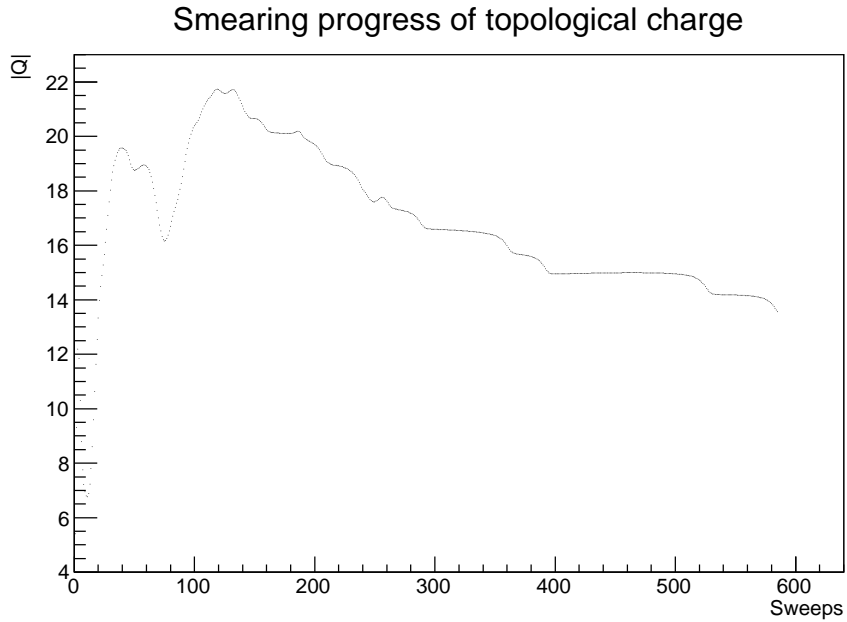}
\caption{\label{fig:q}The development of the topological charge under smearing on a typical configuration of the 24$^4$ lattice for $\beta=2.2$.}
\end{figure}

The primary goal of this work is to understand the behavior of propagators in a topological, i.\ e.\ self-dual, background. That such a background is reached is exemplified in figure \ref{fig:q}. It is visible that at about 300 smearing sweeps the topological charge, even for the very high value of this configuration, has equilibrated and become almost integer\footnote{Note that the charge is always rounded to the next integer in the rest of the text. Especially when changing from one charge level to the next, this is of course rather approximate.}. For smaller charger, this state is usually reached earlier.

\begin{figure}
\includegraphics[width=\linewidth]{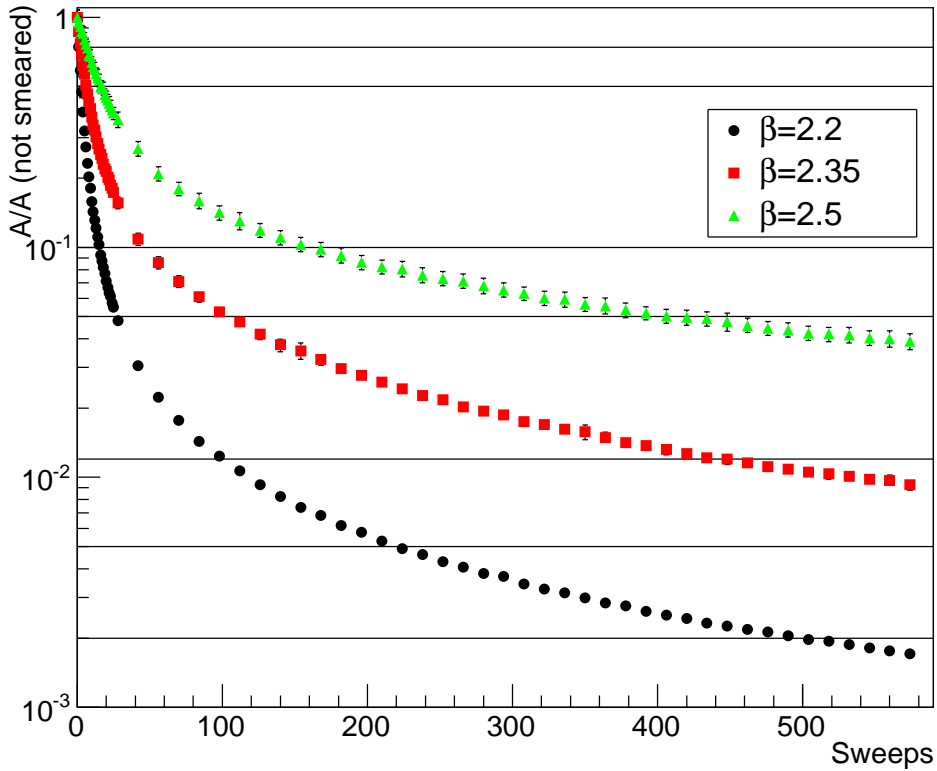}
\caption{\label{fig:ar}The impact of smearing on the interaction measure \pref{alpha} for the 24$^4$ lattices. For details, see text. The horizontal lines are the chosen suppressions for displaying the results in section \ref{sres}.}
\end{figure}

However, in general for different lattice spacings smearing has quite different (quantitative) effects.  The same is also true for different smearing techniques. Hence, a meaningful discussion of the effects requires some comparable quantity \cite{Bonati:2014tqa}. Since the main aim of the present work are the properties of the elementary degrees of freedom, the running coupling appears to be a suitable measure to compare the impact of smearing. Its definition and other properties will be discussed below in section \ref{salpha}. Here, it suffices that it is proportional to some function $A(p)$. Thus, the measure will be given by $A_\text{smeared}(p)/A(p)$, at some fixed $p$. The momentum should be such that $A(p)$ becomes a monotonous function of the smearing sweeps. As will be seen in section \ref{salpha}, this is the case at sufficiently large momenta. Since the different $\beta$ yield a quite different set of momenta, there is no momentum common to all lattices. It is therefore possible to either interpolate the momenta or choose a momentum closest to a reference momentum. Since the results turn out to be only very weakly affected by the choice, here the momentum closest to the reference momentum of 1.65 GeV is chosen. The resulting smearing-dependence is shown in figure \ref{fig:ar}.

As is visible, the decrease is for all values of $\beta$ monotonous, and slower the finer the lattice. This is actually not an effect of the different volumes. The curves are only slightly quantitatively altered by changing the volumes among the available ones, a few-percent-level effect at most. Thus, in the following only the largest volumes will be considered.

It is then visible that the suppression reached for the smearing sweeps required to reach the full self-dual regime according to figure \ref{fig:q} is very large. Such a suppression is not reached on the finer lattices. However, as will be seen later the results show a rather smooth development with the number of smearing sweeps, and the the behavior deep in the self-dual regime differs very little from the one obtained much earlier in the smearing progress.

Hence, the following smearing levels will be used for comparison of the different lattice spacings employed:
\begin{itemize}
 \item The smallest amount of suppression possible for all lattice spacings is 0.74, corresponding to the set of 1, 2, and 6 smearing sweeps for $\beta$ being 2.2, 2.35, and 2.5, respectively
 \item A suppression factor of 2 is reached after 2, 6, and 15 smearing sweeps
 \item A suppression factor of 10 is reached after 15, 42, and 154 smearing sweeps\footnote{As can be seen from table \ref{configs}, for more than 25 smearing sweeps not all intermediate configurations have been gauge-fixed and recorded. Numbers here refer to the situation on a 24$^4$ lattice. On smaller lattices, not always these numbers were available, and thus the closest number has been chosen. This is a negligible effect, of the same order at most as the volume-dependence of the suppression factor.}
 \item A suppression factor of 20 is essentially the upper limit for the finest lattice, reached after 25, 98, and 406 smearing sweeps
 \item A suppression factor of about 85 is the largest reachable on the second-to-finest lattice, giving a 112 and 448 smearing sweeps for $\beta$ values of 2.2 and 2.35, respectively
 \item The onset of self-duality is reached at about a suppression factor of 200 after 210 smearing levels for $\beta=2.2$
 \item A suppression factor of 500 is deep in the self-dual regime, after 490 smearing weeps
\end{itemize} 
The levels of smearing will be used now to compare the results on the different lattices. As noted, however, the development is smooth, and therefore conclusions can be drawn irrespective of where on the smearing trajectories a result is - the development along the trajectory is the truly relevant result.

\section{Results}\label{sres}

The main aim of this work is to understand how the propagators behave for topological configurations. As noted before, this situation is only reached for the coarsest lattices with the available resources. However, as the following will show the changes of the propagators are both monotonous and smooth under smearing. Especially, no qualitative change is observed when entering the regime of pure topological field configurations. It appears thus likely that this is also true for finer lattices.

As a consequence, the main observation of this work is the development under smearing. Hence, in the following the development of the correlation functions under smearing will be presented, and no distinction will be made between the situation at few levels of smearing and the self-dual regime, except where noted, or indicated in the figure. It will be observed that there a trends under smearing, and these trends are stable. In fact, the results in the topological regime differ from those after only few levels of smearing only quantitatively, but not qualitatively. Also this is an important result in itself.

\subsection{Gluon propagator: Momentum space}\label{smom}

\subsubsection{Lattice artifacts}

\begin{figure}
\includegraphics[width=\linewidth]{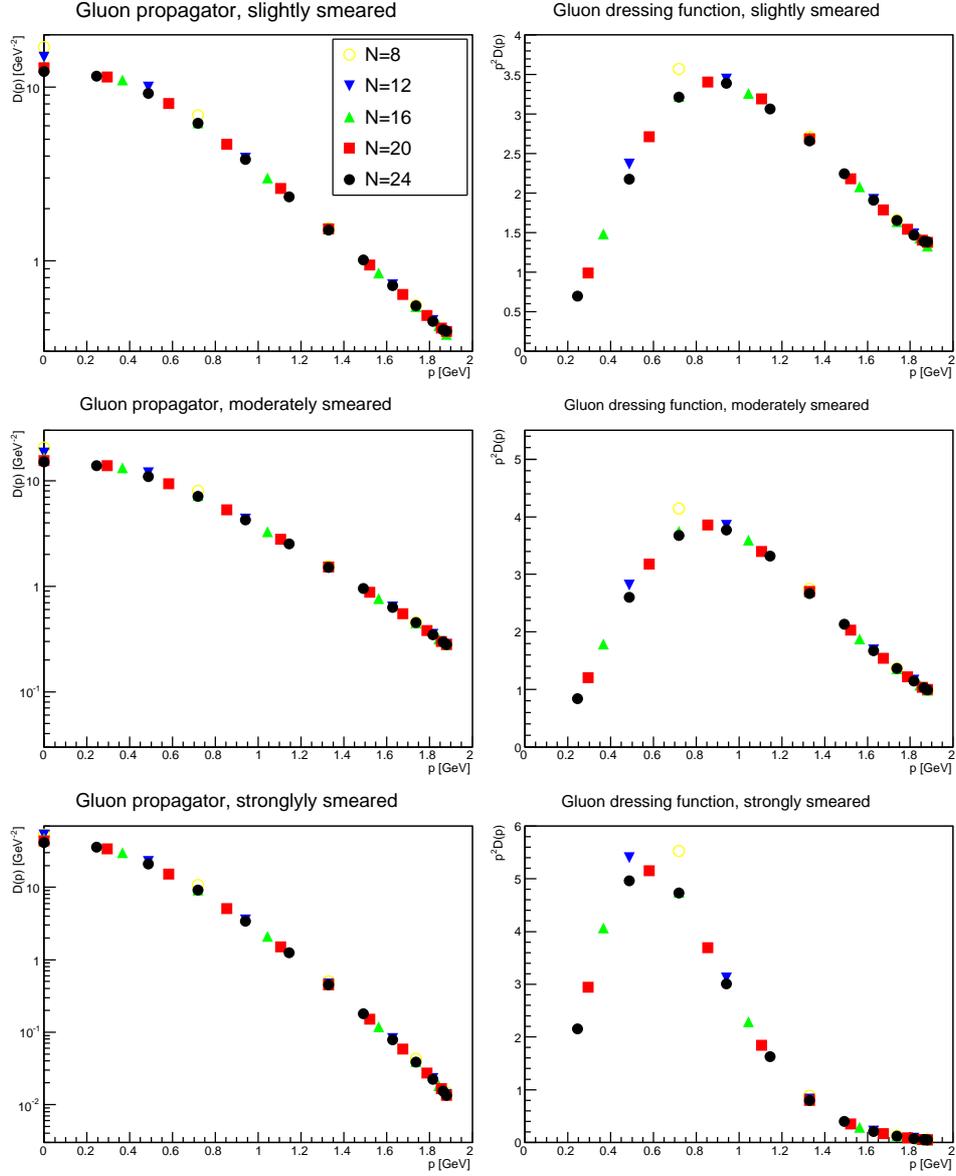}
\caption{\label{fig:v}The gluon propagator (left panels) and dressing function (right panels) at $\beta=2.2$ for different volumes for different numbers of APE sweeps, being slightly smeared (top panels), moderately smeared (middle panels), or strongly smeared (bottom panels). If no errors bars are visible here and hereafter, then they are smaller than the symbol size. The momenta are always along the $x$-axis, best suited to reach low momenta \cite{Maas:2014xma}.}
\end{figure}

The simplest object to investigate is the gluon propagator. To assess the results requires to estimate the two potential types of lattice artifacts, which can affect it, the (physical) volume, and the discretization \cite{Maas:2011se}. Without smearing, the dominant qualitative artifact is the volume \cite{Maas:2011se}, while discretization artifacts yield only a (sizable) quantitative correction, see e.\ g.\ \cite{Bornyakov:2009ug,Oliveira:2012eh,Maas:2014xma}. To assess the effects, the impact of the volume for a slightly (suppression factor 1.35), moderately (suppression factor 2), and strongly (suppression factor 20) smeared gluon propagator is shown in figure \ref{fig:v}. In all cases the extent of smearing is not altering the strength of the finite volume effects significantly.

\begin{figure}
\includegraphics[width=\linewidth]{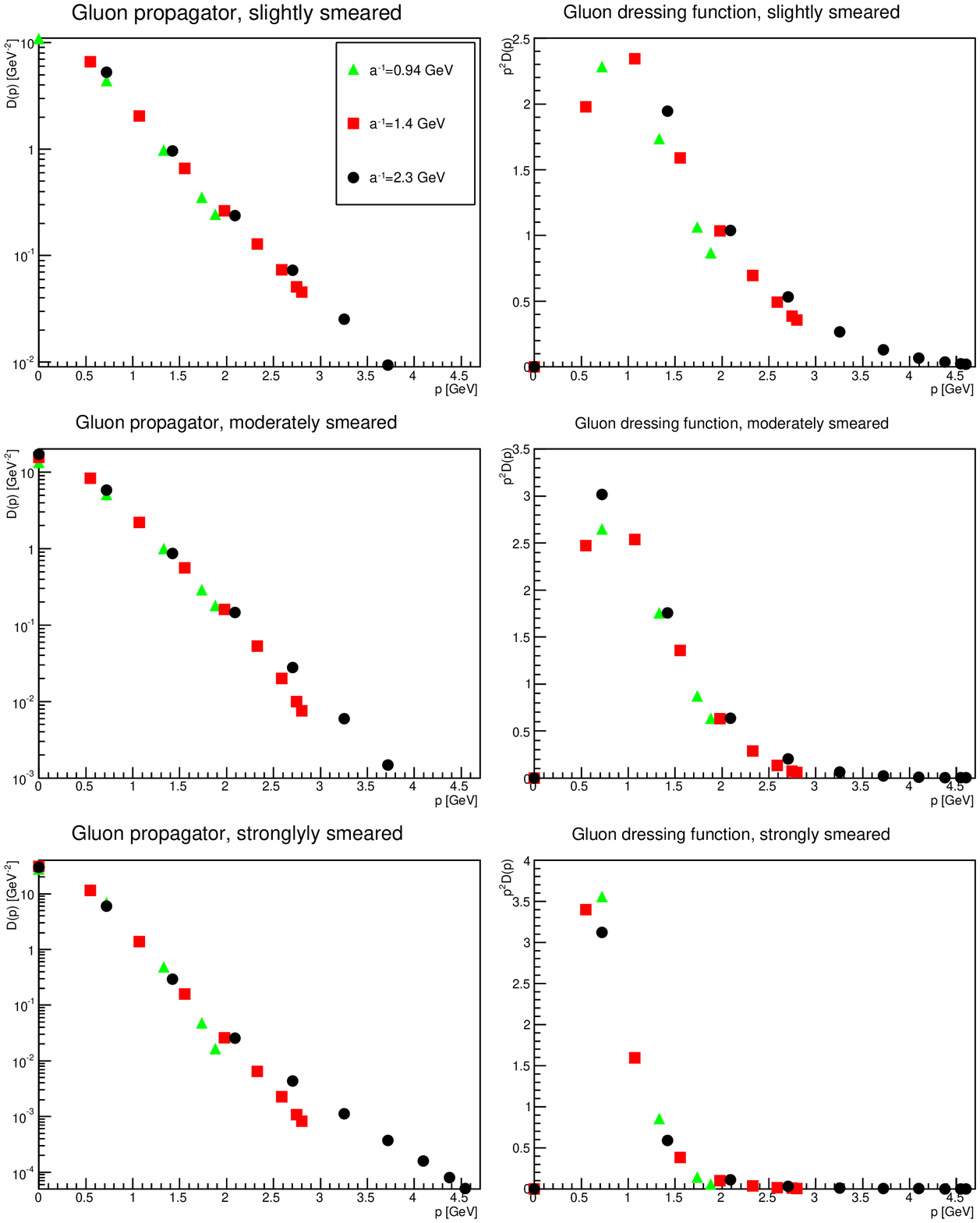}
\caption{\label{fig:a}The gluon propagator (left panels) and dressing function (right panels) for different discretizations at fixed physical volume (1.7 fm)$^4$ for different number of APE sweeps, being slightly smeared (top panels), moderately smeared (middle panels), or strongly smeared (bottom panels). Results are renormalized at 2 GeV.}
\end{figure}

The situation is somewhat different when the discretization is varied, as is visible in figure \ref{fig:a}. While without smearing the ultraviolet part agrees within a few percent for these discretizations \cite{Maas:2014xma}, already slightly smearing changes this. Then, the ultraviolet tail is the stronger suppressed the coarser the discretization. This effect also increases with increasing number of APE sweeps. This is most visible at the renormalization point $\mu=2$ GeV, where the dressing functions coincide without smearing, but differ after strong smearing. The effect is much less pronounced at small momenta. Thus the low-momentum regime is not overmuch affected by discretization effects, but the high-momentum tail is. This is not too surprising, since ultraviolet fluctuations are most affected by the smearing operation, and thus discretization effects should become more pronounced at large momenta.

\subsubsection{Results for the gluon propagator}

\begin{figure}
\includegraphics[width=\linewidth]{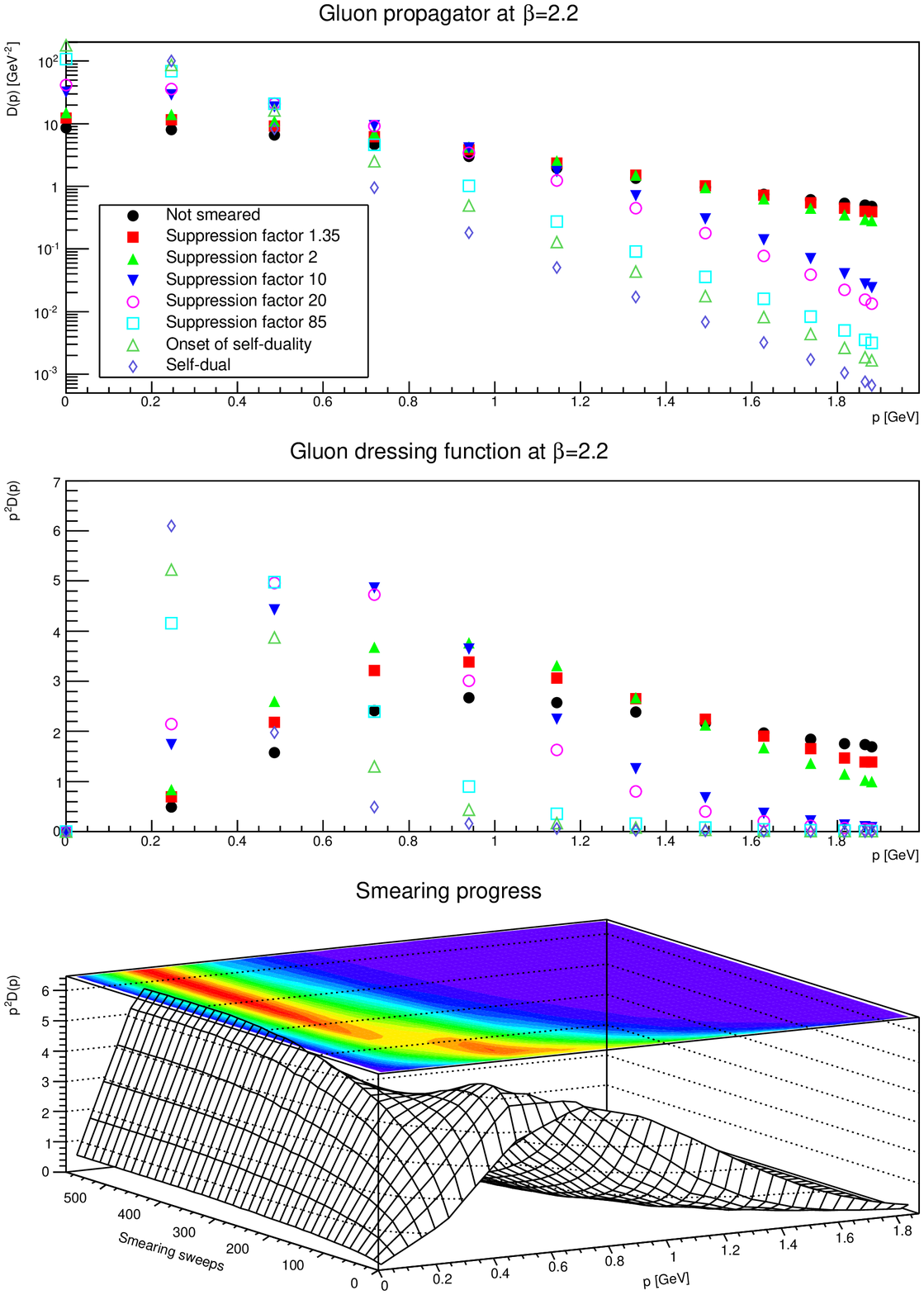}
\caption{\label{fig:gp2}The gluon propagator (top panel) and dressing function (second-to-top panel) for different number of APE sweeps. The plot in the bottom panel shows the dressing function as a function both of momenta and APE sweeps. The discretization is $a=0.21$ fm. All results from $24^4$ lattices.}
\end{figure}

\begin{figure}
\includegraphics[width=\linewidth]{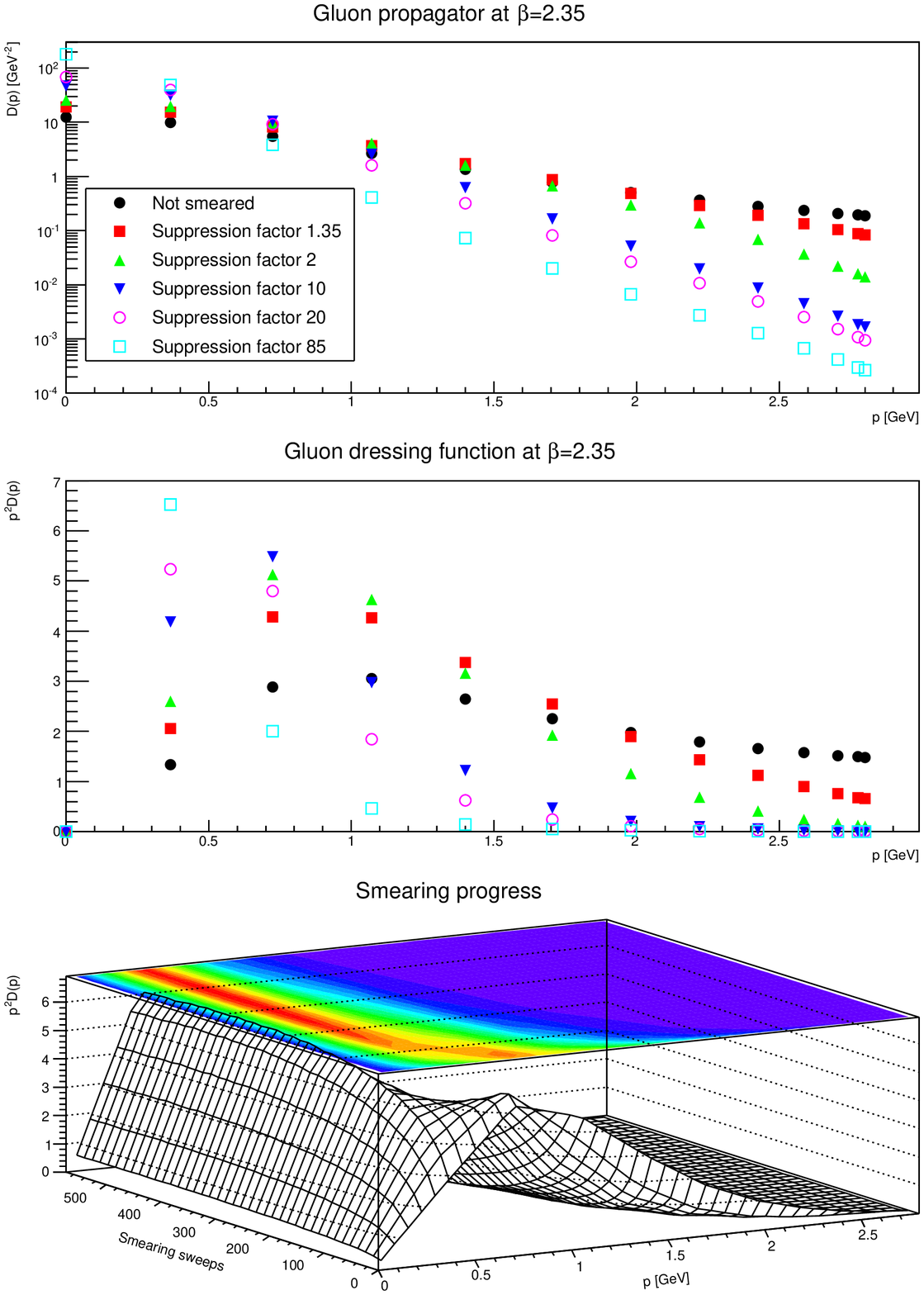}
\caption{\label{fig:gp3}The gluon propagator (top panel) and dressing function (second-to-top panel) for different number of APE sweeps. The plot in the bottom panel shows the dressing function as a function both of momenta and APE sweeps. The discretization is $a=0.14$ fm. All results from $24^4$ lattices.}
\end{figure}

\begin{figure}
\includegraphics[width=\linewidth]{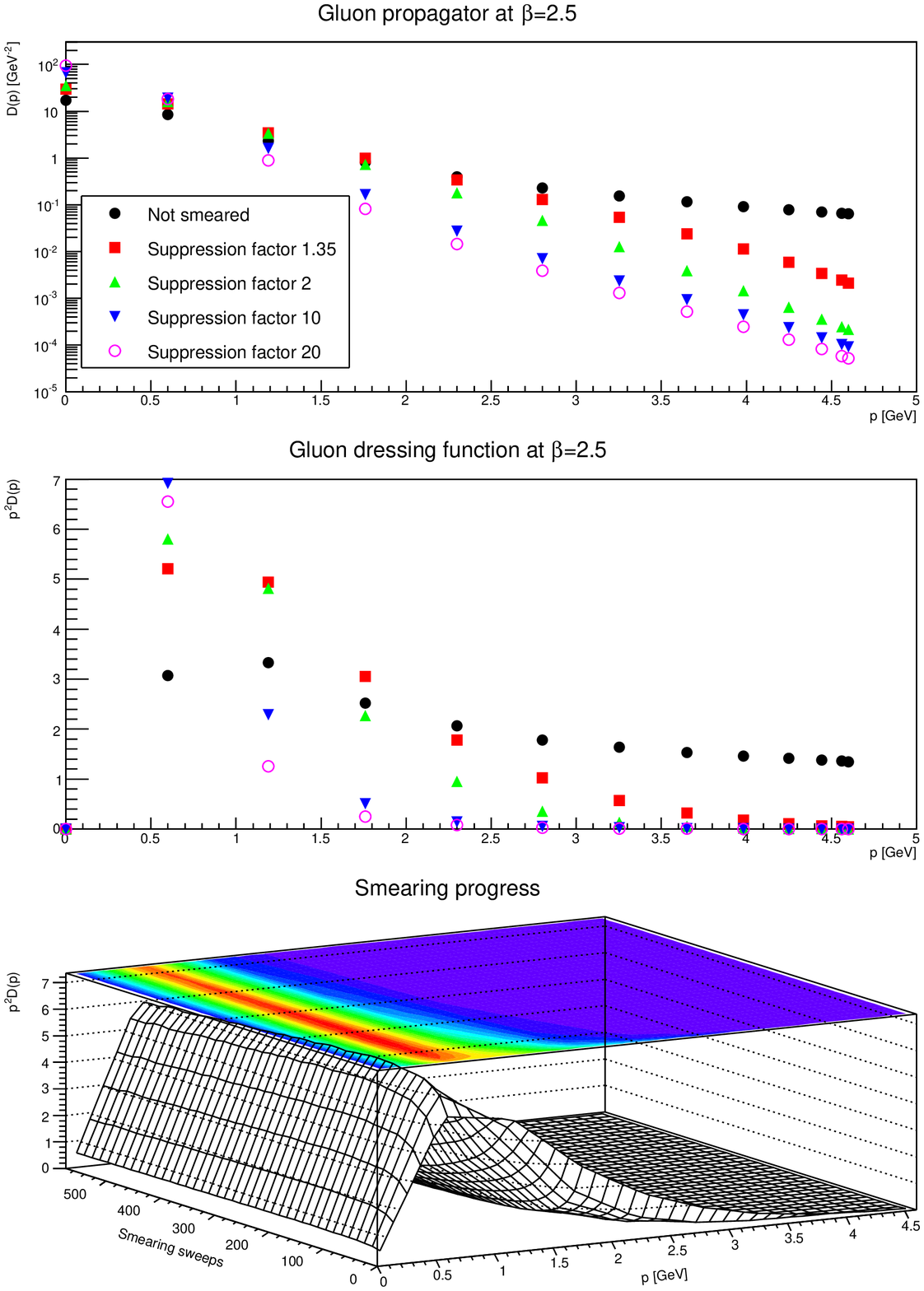}
\caption{\label{fig:gp5}The gluon propagator (top panel) and dressing function (second-to-top panel) for different number of APE sweeps. The plot in the bottom panel shows the dressing function as a function both of momenta and APE sweeps. The discretization is $a=0.087$ fm. All results from $24^4$ lattices.}
\end{figure}

The gluon propagator and dressing functions as a function of the number of smearing sweeps is shown in figures \ref{fig:gp2}-\ref{fig:gp5}. While a suppression by a small factor is not a large effect, even for a rather fine discretization, already moderate suppression factors alter the ultraviolet behavior substantially. This is not surprising, as the ultraviolet tail of the propagators are dominated by the hard, perturbative fluctuations, which are suppressed by the smearing process. It is also visible that with increased smearing the suppression starts at smaller and smaller momenta.

\begin{figure}
\includegraphics[width=\linewidth]{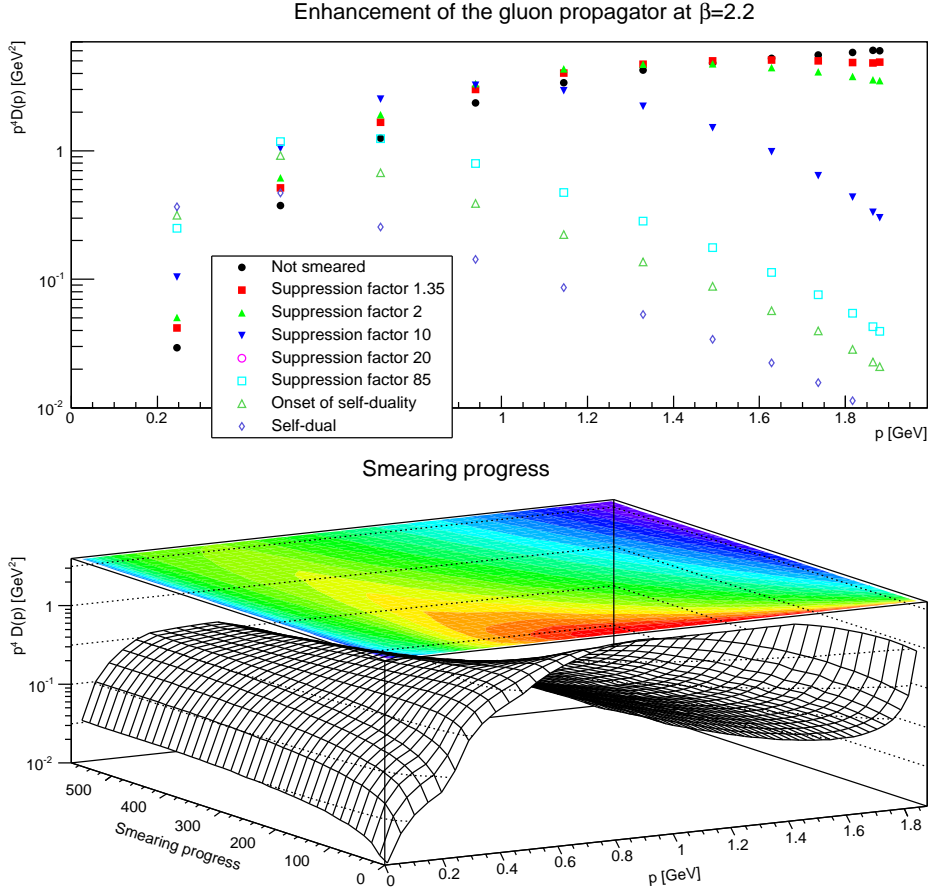}
\caption{\label{fig:gp4}The gluon propagator multiplied by $p^4$ (top panel) for different number of APE sweeps. The plot in the bottom panel shows the dressing function as a function both of momenta and APE sweeps. The discretization is $a=0.21$ fm. All results from $24^4$ lattices.}
\end{figure}

The situation is rather different at small momenta, and shows strong dependence on the physical volume. Concentrating on the case of the largest physical volumes, it is found that the peak of the dressing function becomes enhanced and moves to lower momenta. At the same time, the slope towards the peak becomes increased, appearing to aspire to a $1/p^4$ behavior, as shown explicitly in figure \ref{fig:gp4}. Indeed, after a sufficient number of sweeps, but already before entering the self-dual regime, the maximum in the dressing function has moved to so small momenta that it can almost no longer be observed on the present volumes. The finer lattices show the same behavior, but, due to the smaller volumes, the peaks disappear much earlier from sight.

Of course, the lattice volumes are small, and therefore this may change once more when moving towards very large volumes. Especially, it cannot be concluded that the peak will survive forever at smaller and smaller momenta. Nonetheless, it appears that the salient feature of the gluon propagator survives under smearing. The appearance of a $1/p^{4}$ behavior at typical hadronic scales is, however, interesting. Especially, as such a strong rise coincides with the naive expectation for a linear rising potential \cite{Alkofer:2000wg}. This will be discussed further in section \ref{sspec}.

\begin{figure}
\includegraphics[width=\linewidth]{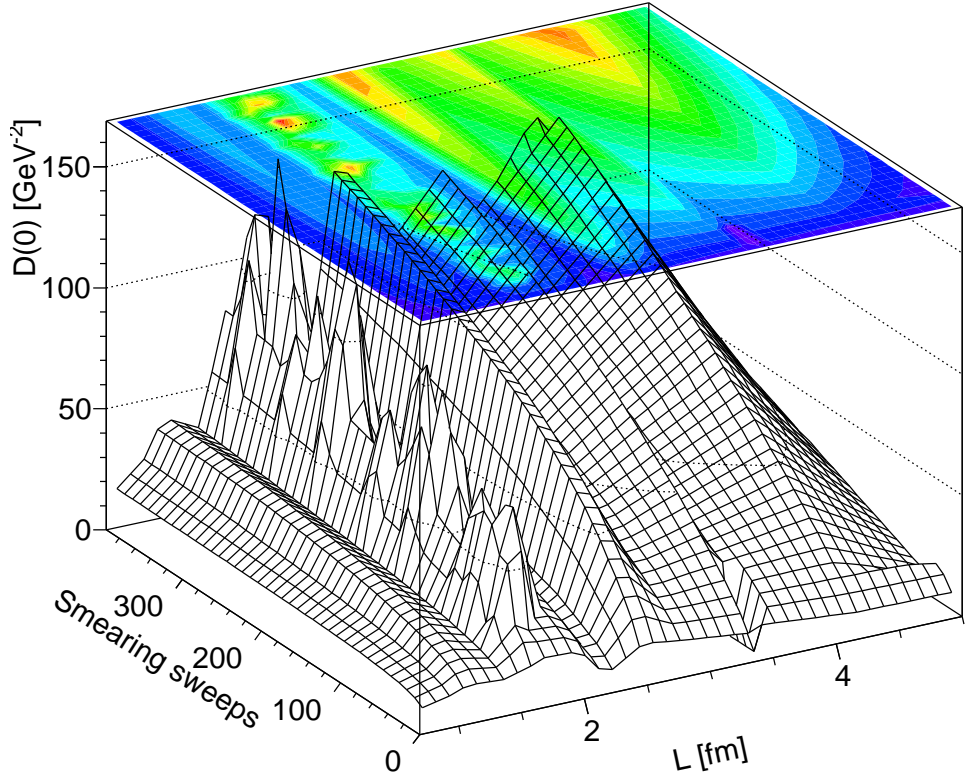}
\caption{\label{fig:d0}The gluon propagator at zero momentum as a function of volume and smearing steps. If for a volume two discretizations are available, both have been included, what especially at small volumes leads to fluctuations. The ridge structure appears due to the substantial impact of various systematic effects at such small volumes and coarse lattices; the relevant information here is merely the common trend. All results are renormalized, as discussed in section \ref{stech}.}
\end{figure}

To trace out this behavior further, a helpful quantity is $D(0)$ as a function of physical volume, which is shown in figure \ref{fig:d0}. Since $D(0)$ more or less continuously increases with the number of smearing steps for all investigated volumes, except perhaps for the largest one, there is not yet any conclusive hint of the fate of the maximum in the dressing function in a topological background field. However, even the largest volumes included have to be considered small with respect to the asymptotic behavior of the gluon propagator \cite{Maas:2011se}, and therefore caution is mandatory.

\subsubsection{Dependency on the topological charge}

Given the observation \cite{DelDebbio:2002xa} that lattice simulations algorithms tend to become stuck in a sector of fixed (net) topological charge when moving further and further towards the continuum limit, it is a relevant question whether this affects the correlation functions substantially, so that special care would be required. In general, several APE sweeps are necessary before the topological charge stabilizes itself, so to answer this question for the unsmeared case requires to extrapolate any dependency backwards. One substantial problem in doing so is that the topological charge is Gaussian distributed for configurations, and therefore very large statistics is needed to access large topological charge, especially for highly smeared configurations.

\afterpage{\clearpage}

\begin{figure}
\includegraphics[width=\linewidth]{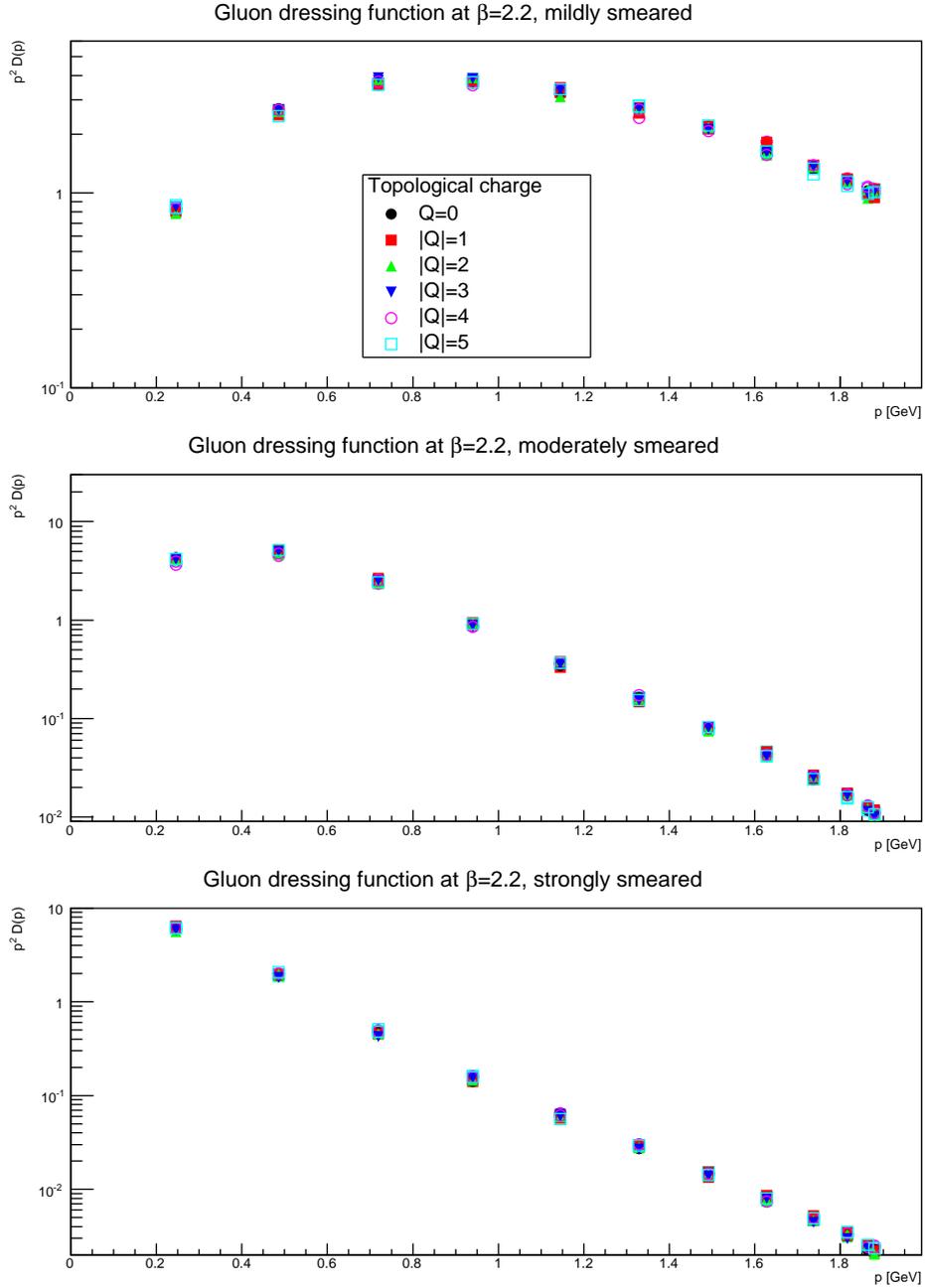}
\caption{\label{fig:gp-q2}The gluon dressing function for mildly (suppression factor 2, top panels), moderately (suppression factor 85, middle panels), and strongly smeared (self-dual regime, bottom panels) configurations, in different fixed topological charge sectors. The lattice spacing is $a=0.21$ fm, on a $24^4$ lattices.}
\end{figure}

\begin{figure}
\includegraphics[width=\linewidth]{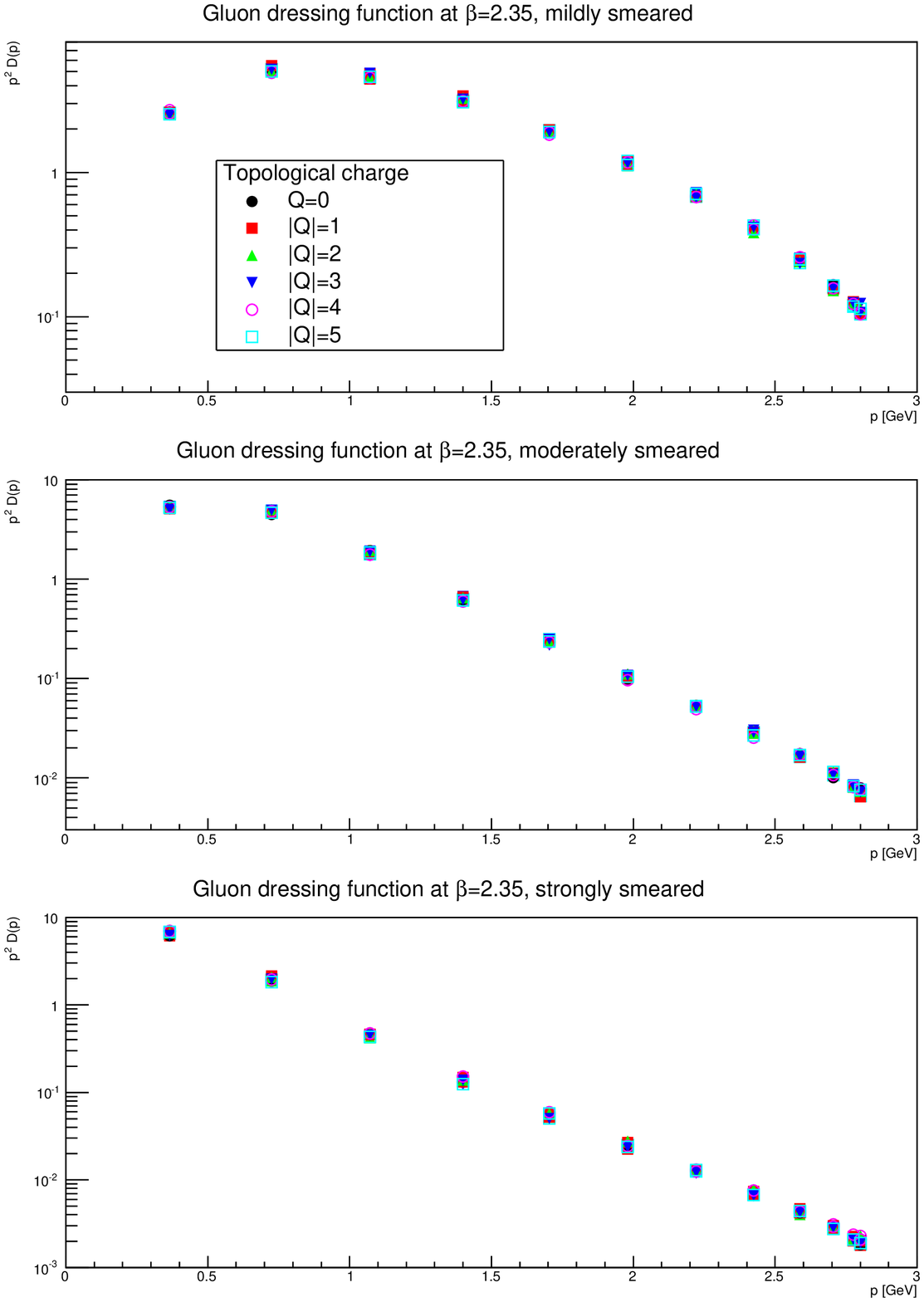}
\caption{\label{fig:gp-q3}The gluon dressing function for mildly (suppression factor 2, top panels), moderately (suppression factor 20, middle panels), and strongly smeared (suppression factor 85 bottom panels) configurations, in different fixed topological charge sectors. The lattice spacing is $a=0.14$ fm, on a $24^4$ lattices.}
\end{figure}

\begin{figure}
\includegraphics[width=\linewidth]{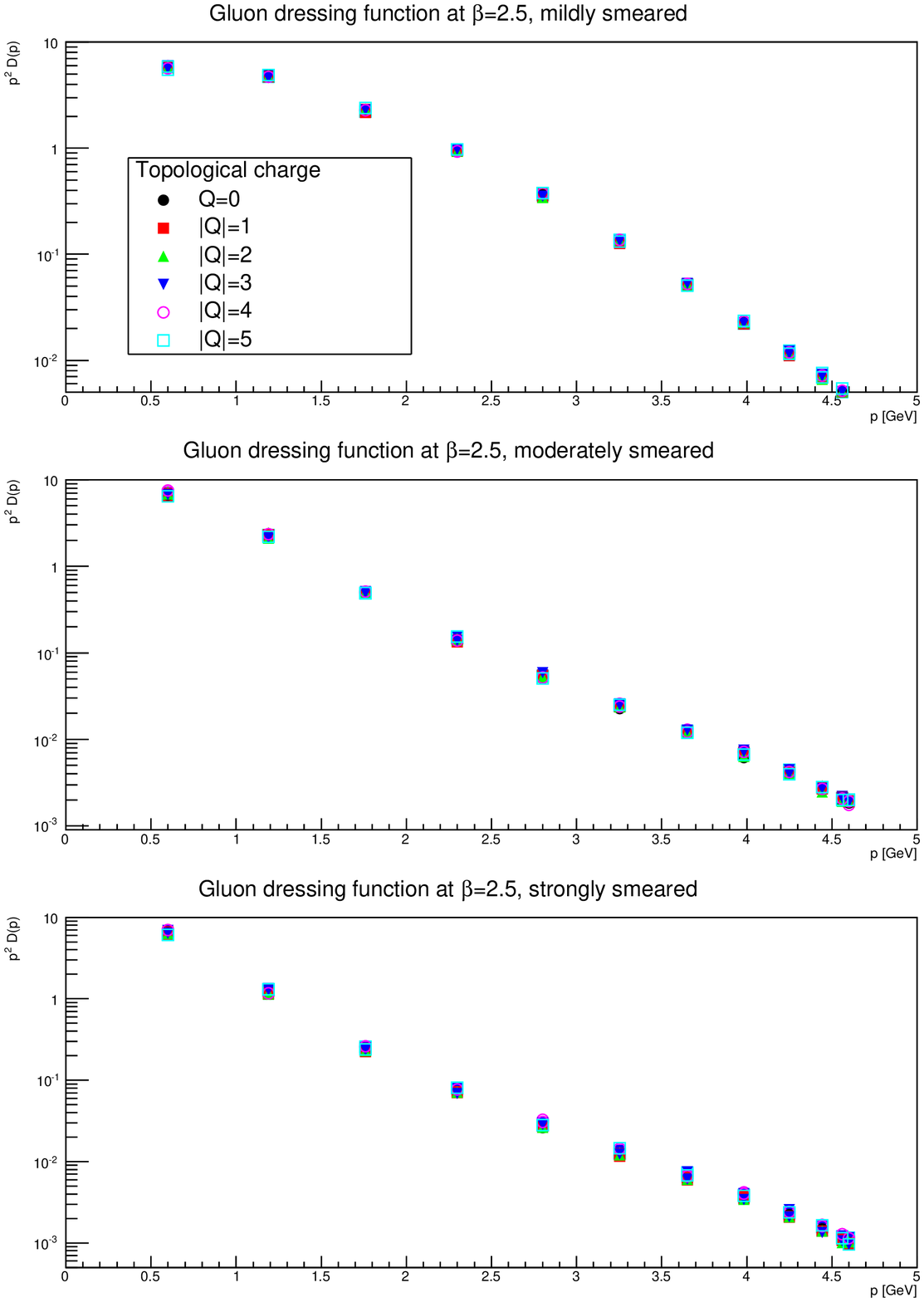}
\caption{\label{fig:gp-q5}The gluon dressing function for mildly (suppression factor 2, top panels), moderately (suppression factor 10, middle panels), and strongly smeared (suppression factor 20, bottom panels) configurations, in different fixed topological charge sectors. The lattice spacing is $a=0.087$ fm, on a $24^4$ lattices.}
\end{figure}

\afterpage{\clearpage}

Keeping this statistical limitation in mind, results for several discretizations and smearing sweeps are shown in figure \ref{fig:gp-q2}-\ref{fig:gp-q5}. It is visible that there is essentially no effect, no matter the suppression factor nor whether in the self-dual domain or not. Thus, the gluon propagator does not appear to depend too strongly on the topological charge sector.

It should again be stressed that $Q$ is a net-charge. Therefore, this result cannot characterize how the propagator would look, e.\ g., in configurations with two lumps of topological charge 1 and one with -1 in contrast to a configuration with only one of charge 1. The dependence on such a characterization may be quite different than the one found here.

\subsection{Gluon propagator: Position space}\label{spos}

The gluon propagator in position space has been a valuable quantity to indirectly determine generic properties of the the analytic structure of the propagator \cite{Fischer:2006ub,Maas:2011se}. Though by now first direct continuum calculations using functional methods are available for it \cite{Strauss:2012as}, the systematic uncertainties make indirect information still valuable. Since especially the long-range structure is interesting, the smearing could be expected to improve the quality of the results, similar to what is obtained for bound states \cite{Gattringer:2010zz}.

\begin{figure}
\includegraphics[width=\linewidth]{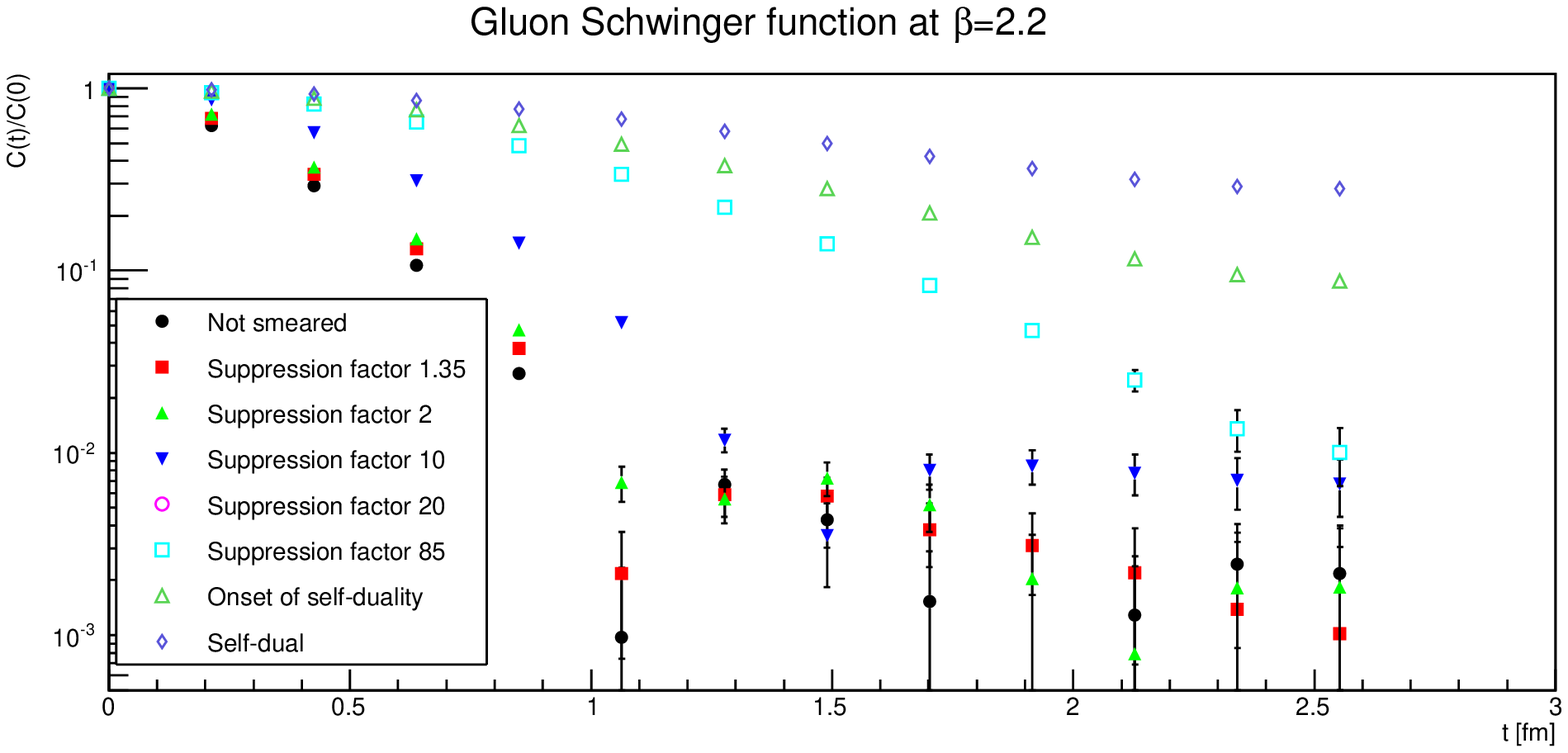}\\
\includegraphics[width=\linewidth]{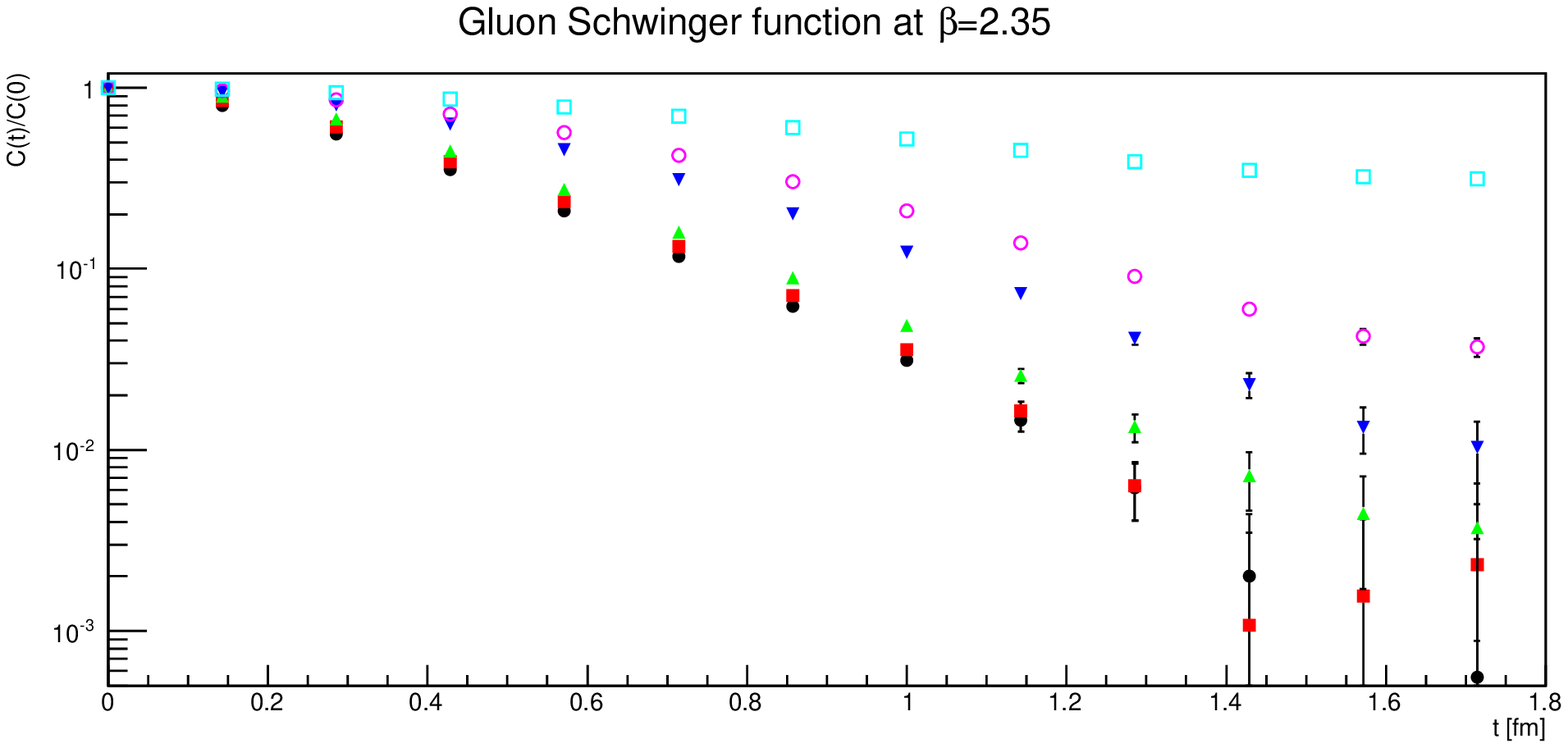}\\
\includegraphics[width=\linewidth]{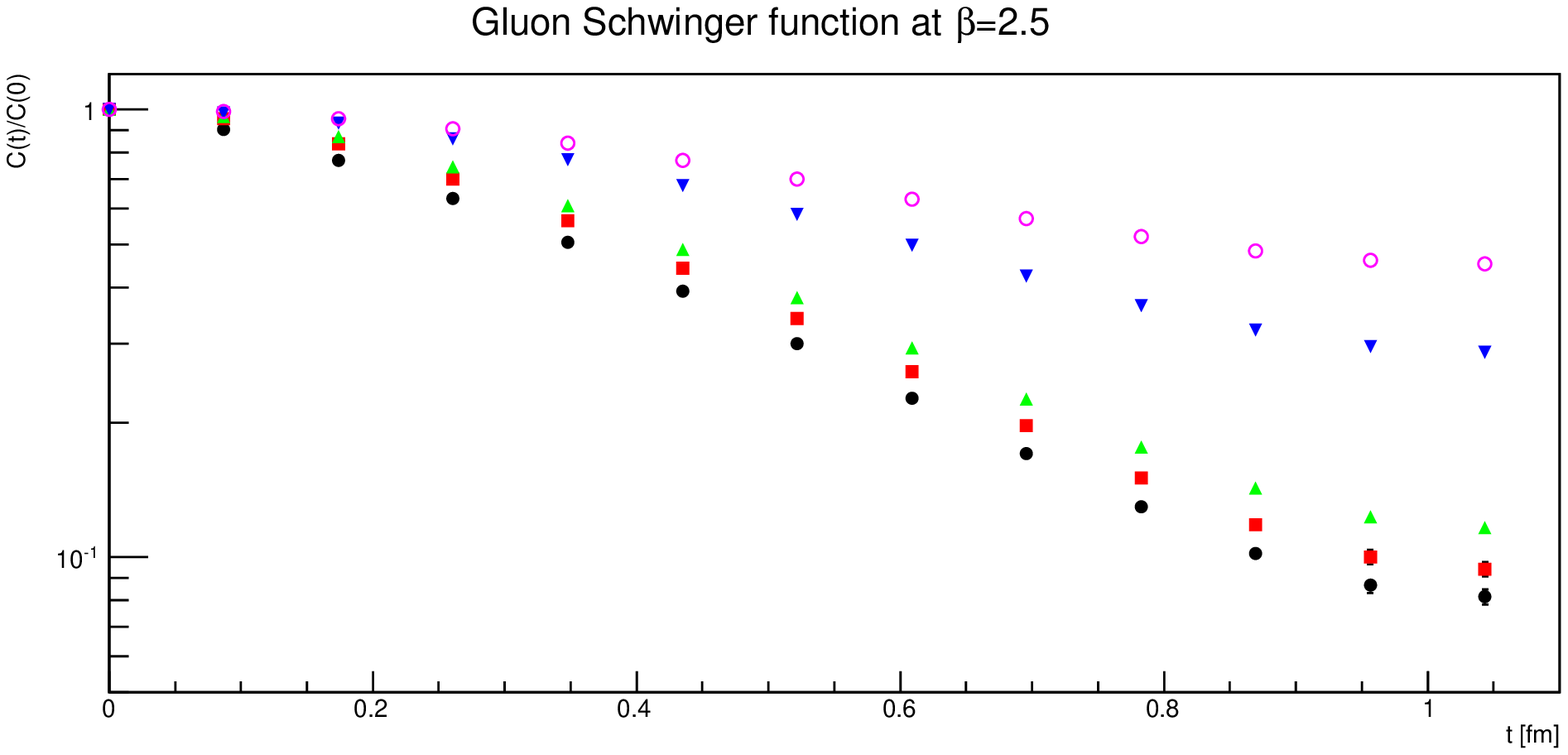}
\caption{\label{fig:gp-s}The gluon dressing function in position space for various numbers of smearing sweeps. The top panel is for a discretization of $a=0.21$ fm, the middle panel for $a=0.14$ fm, and the bottom panel for $a=0.087$ fm. All results from $24^4$ lattices.}
\end{figure}

\afterpage{\clearpage}

The result for different smearing levels are shown for the three discretizations in figure \ref{fig:gp-s}. Some qualitative difference is observed for the different discretizations. This is mainly that the finer the discretization, the earlier a behavior is seen which appears on coarser lattice only for a stronger suppression. This effect is of the same size as the difference due to the different volumes of the not smeared case, and hence is probably rather a finite-volume artifact. It is therefore possible to concentrate on the results for the largest physical volumes, and therefore longest accessible times. It is found that the decay becomes slower the more smearing has been performed. This moves the characteristic zero crossing \cite{Maas:2011se} to larger times, but it is even after substantial smearing still observable. Even in the cases where the zero crossing is not visible, and even after the longest amount of smearing, the correlator still curves incorrectly for a physical correlator. Thus, the generic properties of the gluon propagator remain even in a pure self-dual/topological background.

Of course, if one is willing to interpret the severely smeared configuration as the relevant structure of the configurations, this could also be interpreted as that positivity violations in the gluon propagator, and thus its absence from the physical state spectrum, is caused by topological effects. This interpretation would also imply that the presence of positivity violation can be taken as a sign of incorporating collective gluonic effects, especially when comparing to the results on the residual configurations, discussed below in section \ref{sresi}. However, such an interpretation is always threatened by a possible reversal of cause and symptom.

\subsection{Ghost propagator}\label{sghost}

\begin{figure}
\includegraphics[width=0.5\linewidth]{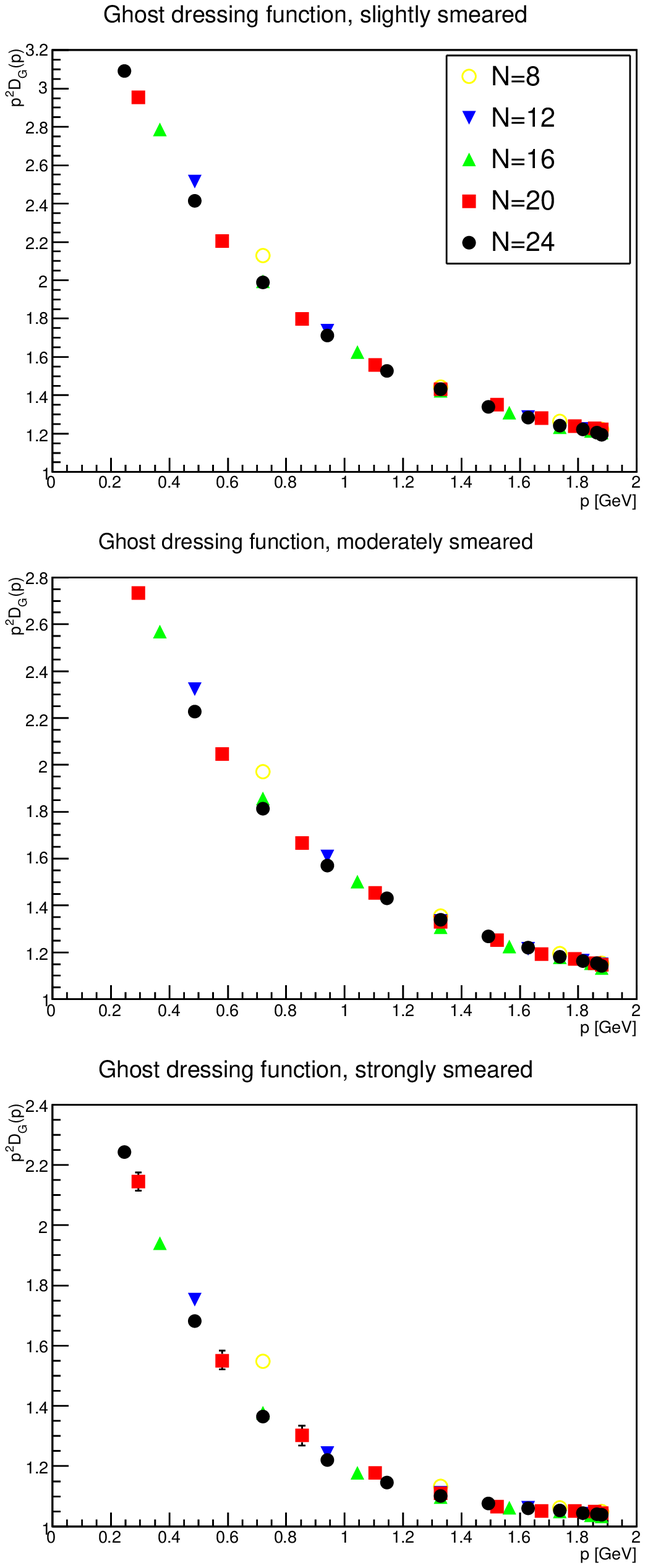}\includegraphics[width=0.5\linewidth]{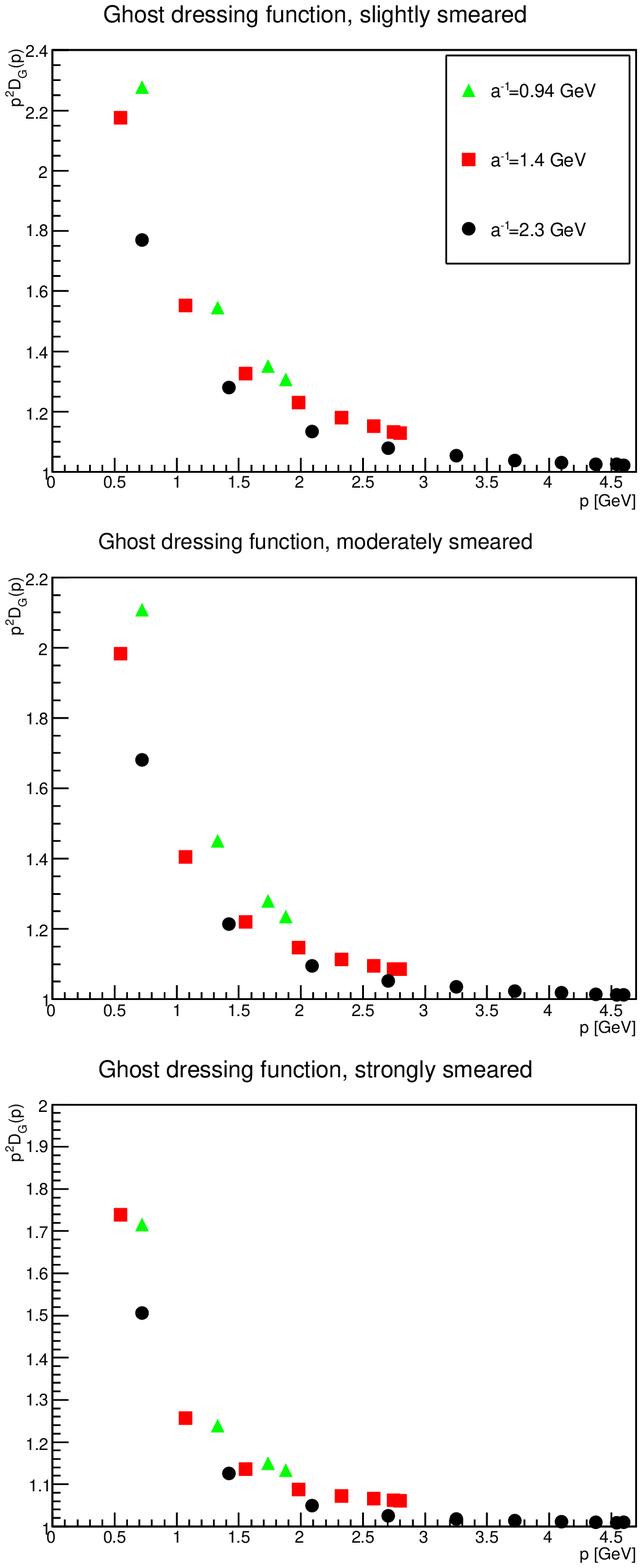}
\caption{\label{fig:la-ghp}The ghost dressing function for different volumes at fixed lattice spacing (left panels) and different lattice spacing at fixed physical volumes (1.7 fm)$^4$ (right panels) for different number of APE sweeps, being slightly smeared, moderately smeared, or strongly smeared.}
\end{figure}

Besides the gluon propagator, the ghost propagator plays an important role as it contributes to the infrared dynamics in Landau gauge, for both Yang-Mills theory and QCD \cite{Maas:2011se,Fischer:2006ub}. Especially, it will be important for the running coupling in the next section \cite{Alkofer:2000wg}. Before analyzing it, the first step is once more to assess the importance of lattice artifacts. The influence of both the lattice volume as well as discretization for a smeared ghost dressing function is shown in figure \ref{fig:la-ghp}, for the same suppression factors as in figure \ref{fig:v} and \ref{fig:a}. In comparison to the effects on the gluon dressing function in figures \ref{fig:v} and \ref{fig:a}, the impact for the ghost dressing function is different. The volume artifacts are very similar to the case without smearing \cite{Maas:2011se}. Changing the discretization has more effect.  One is on the renormalization, which is performed in figure \ref{fig:la-ghp} with the same renormalization constants as for the unsmeared case. These factors yield even for the smallest suppression no longer coinciding values. At the strongest suppression, the effect is more pronounced as it suppressed the ghost dressing function for the coarsest lattice stronger than it is on the finest lattice. Hence, the results discussed below may be bigger on a finer lattice. Also comparing the values of $\beta=2.35$ and $\beta=2.5$, they move closer together in the far infrared with increasing smearing. Thus, the impact of smearing seems to increase with increasing lattice spacing.

\begin{figure}
\includegraphics[width=\linewidth]{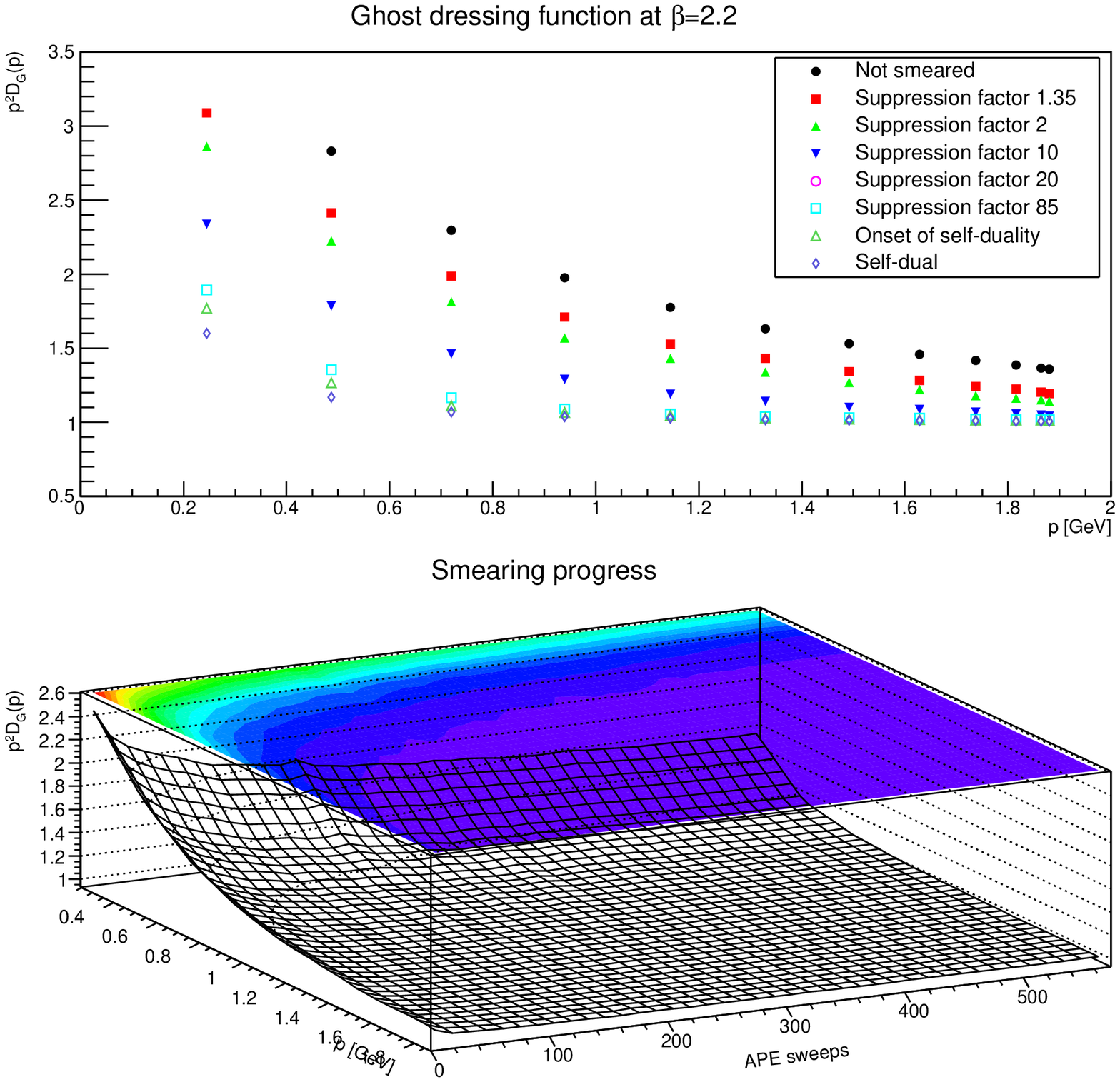}\
\caption{\label{fig:ghp2}The (unrenormalized) ghost dressing function (top panel) for different numbers of APE sweeps. The plot in the bottom panel shows the dressing function as a function both of momenta and APE sweeps.  discretization is $a=0.21$ fm. All results from $24^4$ lattices.}
\end{figure}

\begin{figure}
\includegraphics[width=\linewidth]{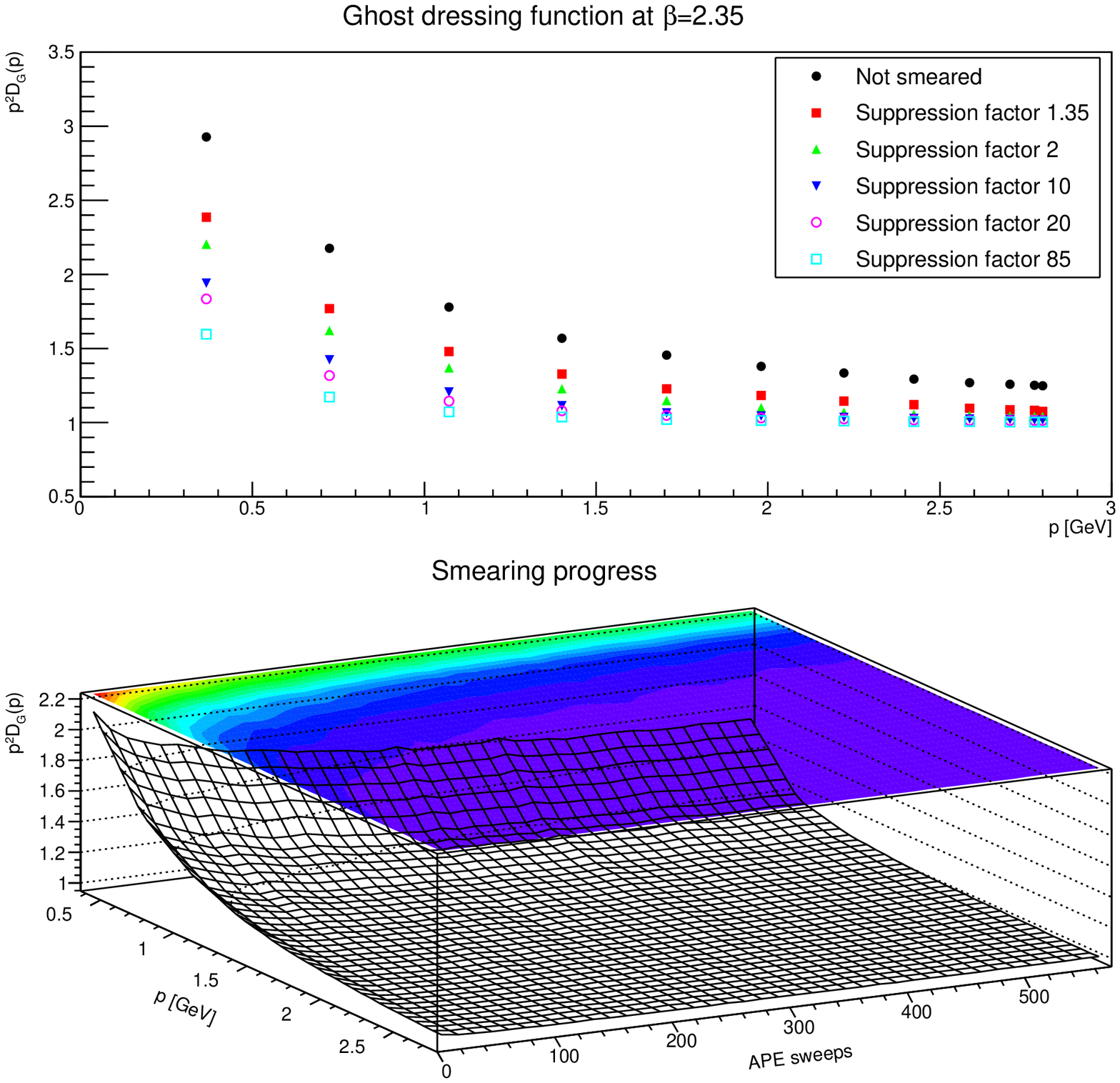}\
\caption{\label{fig:ghp3}The (unrenormalized) ghost dressing function (top panel) for different numbers of APE sweeps. The plot in the bottom panel shows the dressing function as a function both of momenta and APE sweeps.  discretization is $a=0.14$ fm. All results from $24^4$ lattices.}
\end{figure}

\begin{figure}
\includegraphics[width=\linewidth]{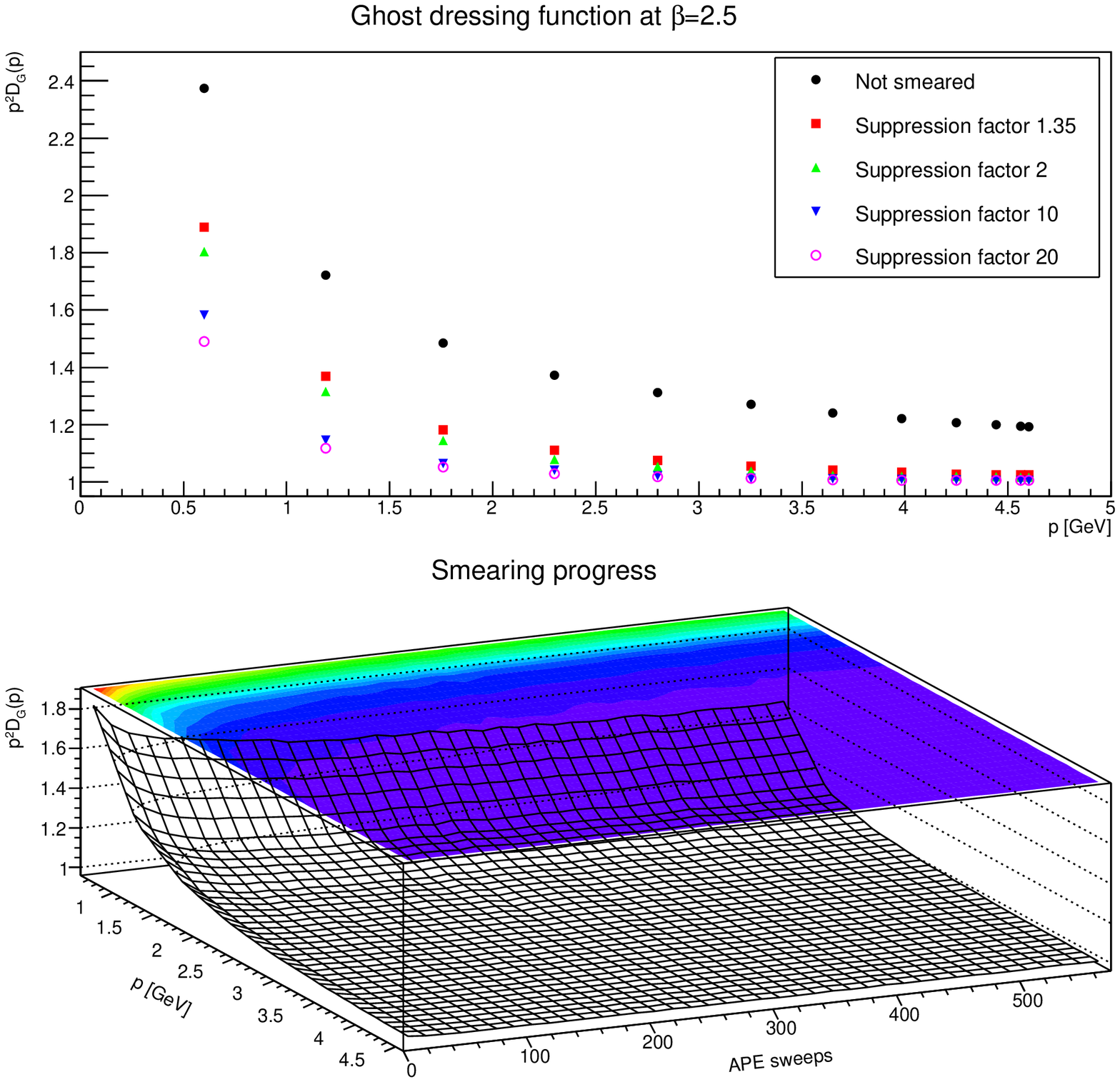}\
\caption{\label{fig:ghp5}The (unrenormalized) ghost dressing function (top panel) for different numbers of APE sweeps. The plot in the bottom panel shows the dressing function as a function both of momenta and APE sweeps.  discretization is $a=0.087$ fm. All results from $24^4$ lattices.}
\end{figure}

Keeping this in mind, the next step is to proceed to the dependence on the smearing progress. The equivalent to figures \ref{fig:gp2}-\ref{fig:gp5} for the gluon is shown for the ghost in figures \ref{fig:ghp2}-\ref{fig:ghp5}. Shown is the unrenormalized ghost dressing function, and that for a particular reason: When increasing the number of APE sweeps, the high-momentum behavior is not removed, like for the gluon propagator. Rather, the behavior of the propagator becomes more and more the one of a free particle, with trivial renormalization. At the same time the qualitative infrared behavior is unaltered, and the ghost dressing function remains enhanced. However, the amount of enhancement is significantly reduced at fixed momenta. The same effect would occur, if the enhancement would be shifted to smaller momenta. Hence, this behavior is in agreement with what is observed for the gluon propagator, where its characteristic infrared behavior sets in for smaller and smaller momenta the more the configurations are smeared, and the main impact occurs at mid-momentum. There is no distinguishing effect when reaching the self-dual domain: The reduction of infrared strength is a continuous process.

\begin{figure}
\includegraphics[width=\linewidth]{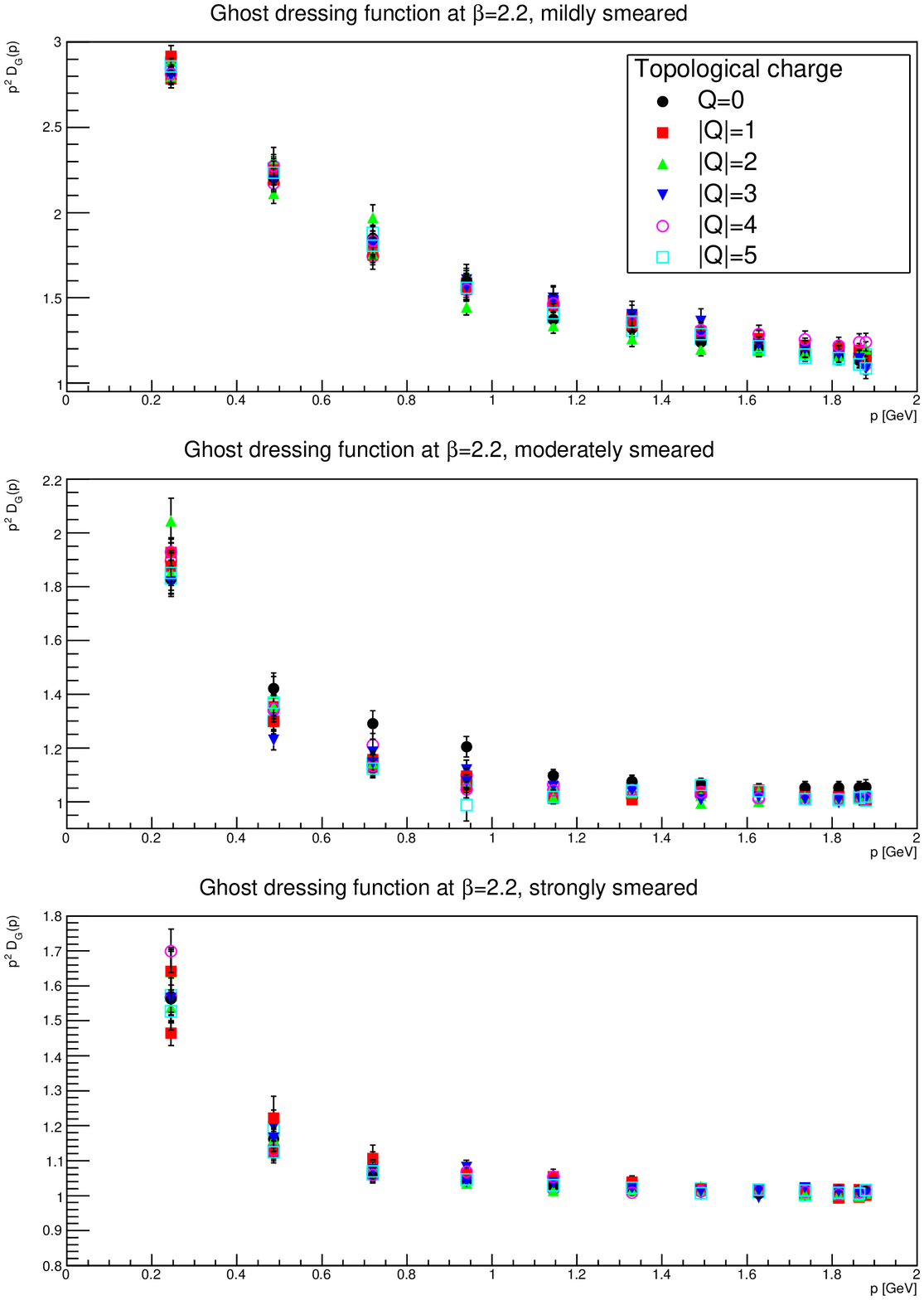}
\caption{\label{fig:ghp-q2}The ghost dressing function for mildly (suppression factor 2, top panels), moderately (suppression factor 85, middle panels), and strongly smeared (self-dual regime, bottom panels) configurations, in different fixed topological charge sectors. The discretization is $a=0.21$ fm. All results from $24^4$ lattices.}
\end{figure}

\begin{figure}
\includegraphics[width=\linewidth]{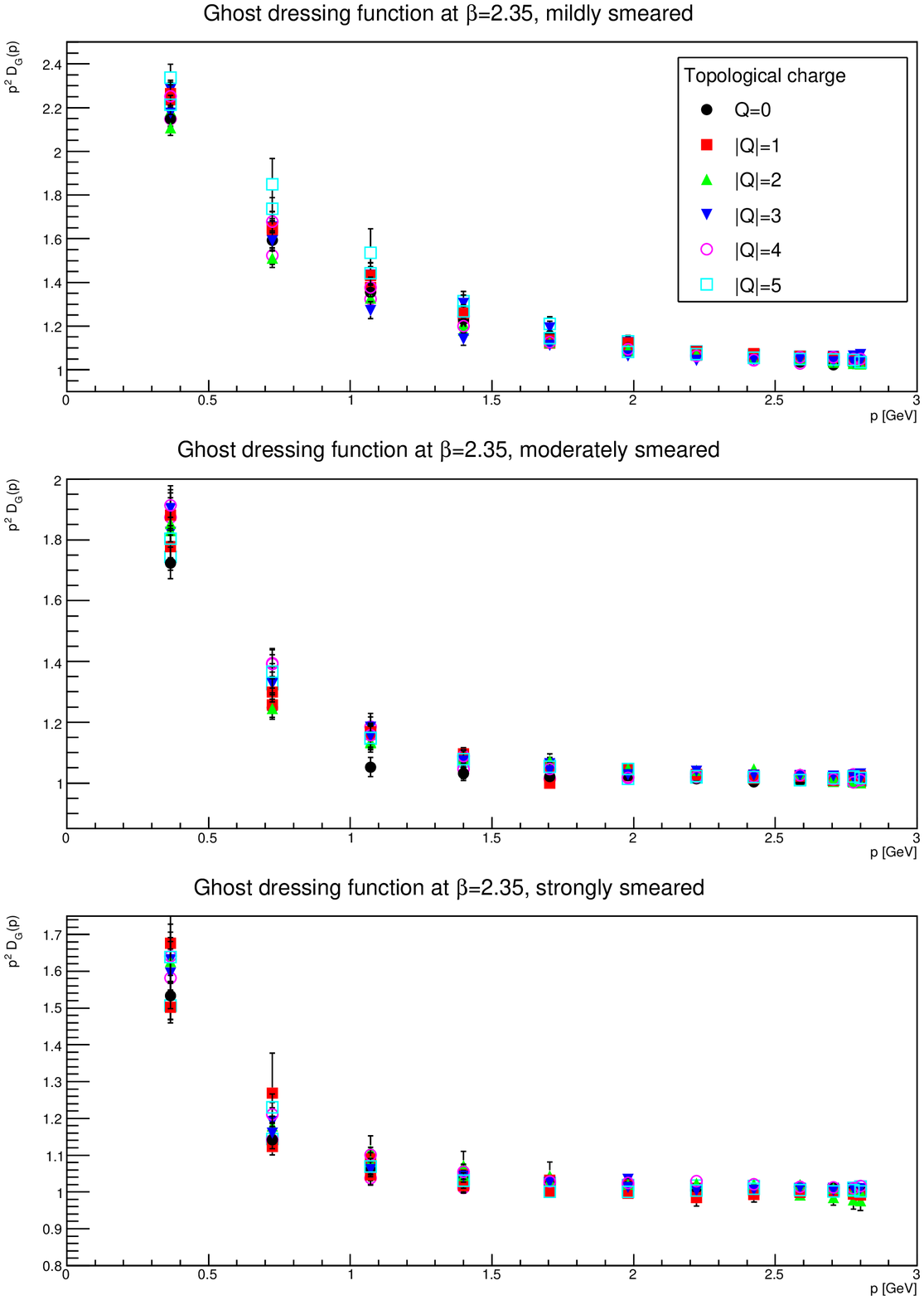}
\caption{\label{fig:ghp-q3}The ghost dressing function for mildly (suppression factor 2, top panels), moderately (suppression factor 20, middle panels), and strongly smeared (suppression factor 85 bottom panels) configurations, in different fixed topological charge sectors. The discretization is $a=0.14$ fm. All results from $24^4$ lattices.}
\end{figure}

\begin{figure}
\includegraphics[width=\linewidth]{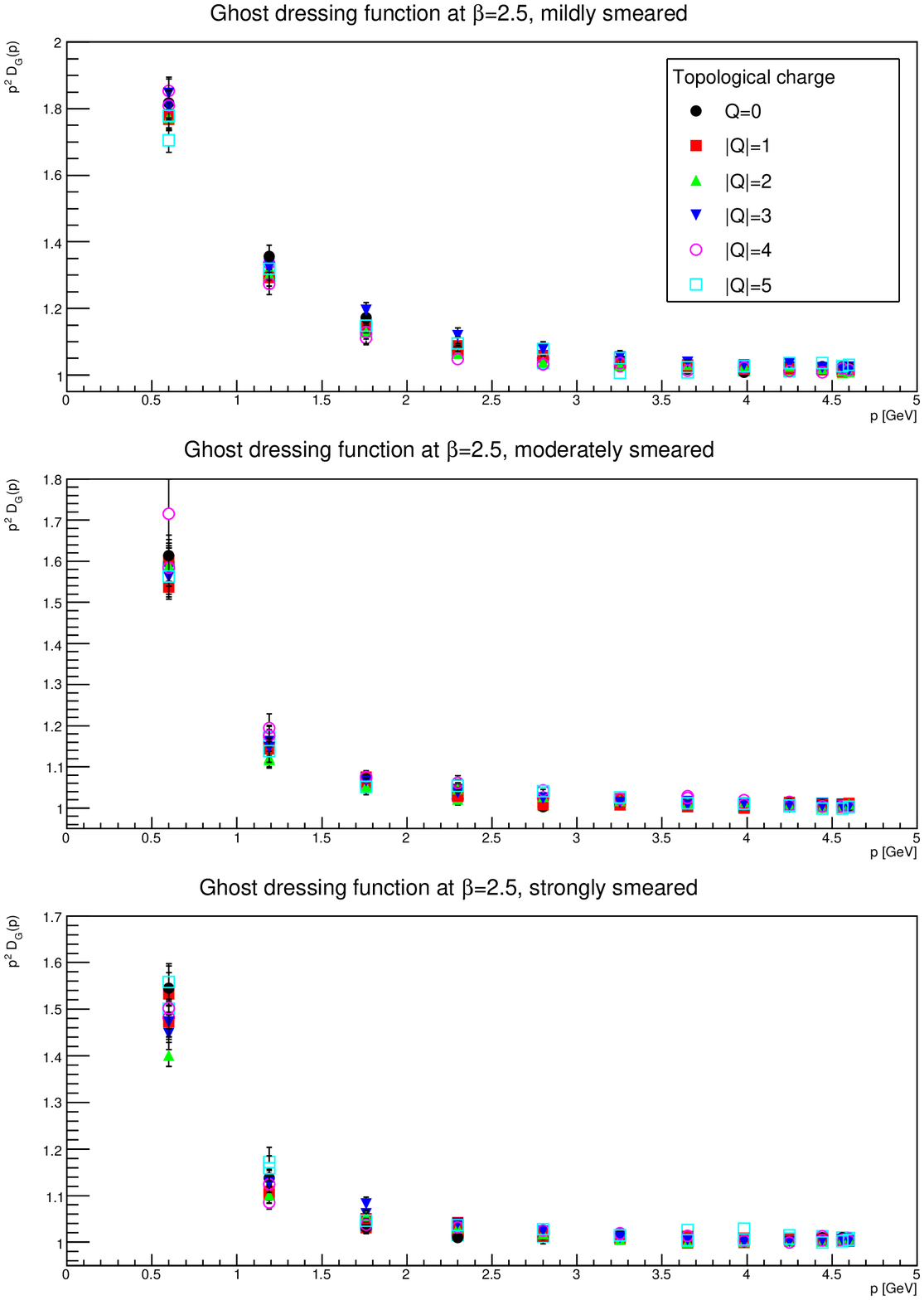}
\caption{\label{fig:ghp-q5}The ghost dressing function for mildly (suppression factor 2, top panels), moderately (suppression factor 10, middle panels), and strongly smeared (suppression factor 20, bottom panels) configurations, in different fixed topological charge sectors. The discretization is $a=0.087$ fm. All results from $24^4$ lattices.}
\end{figure}

For the same reason as for the gluon propagator, is it interesting to identify the topological-(net-)charge dependence of the ghost propagator, which is shown in figure \ref{fig:ghp-q2}-\ref{fig:ghp-q5}. Similarly to the gluon case, no statistically significant dependency on the topological charge is observed. Therefore, the same conclusion holds as for the gluon propagator, i.\ e.\ there is no significant dependency on the topological-(net-)charge sector.

Of course, the same caveat that this is a dependence on a net-charge rather than the number of topological lumps applies here as well.

\subsection{Running coupling}\label{salpha}

\begin{figure}
\includegraphics[width=\linewidth]{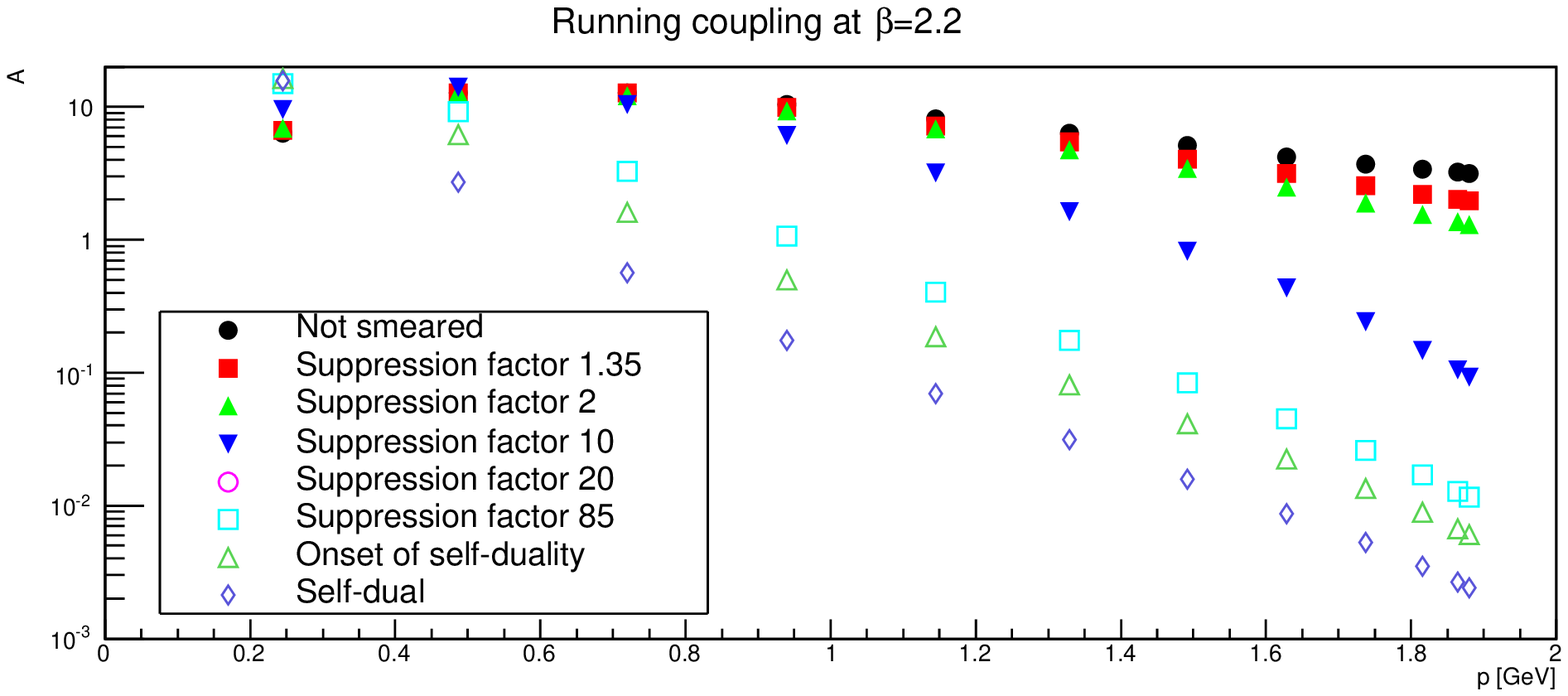}\\
\includegraphics[width=\linewidth]{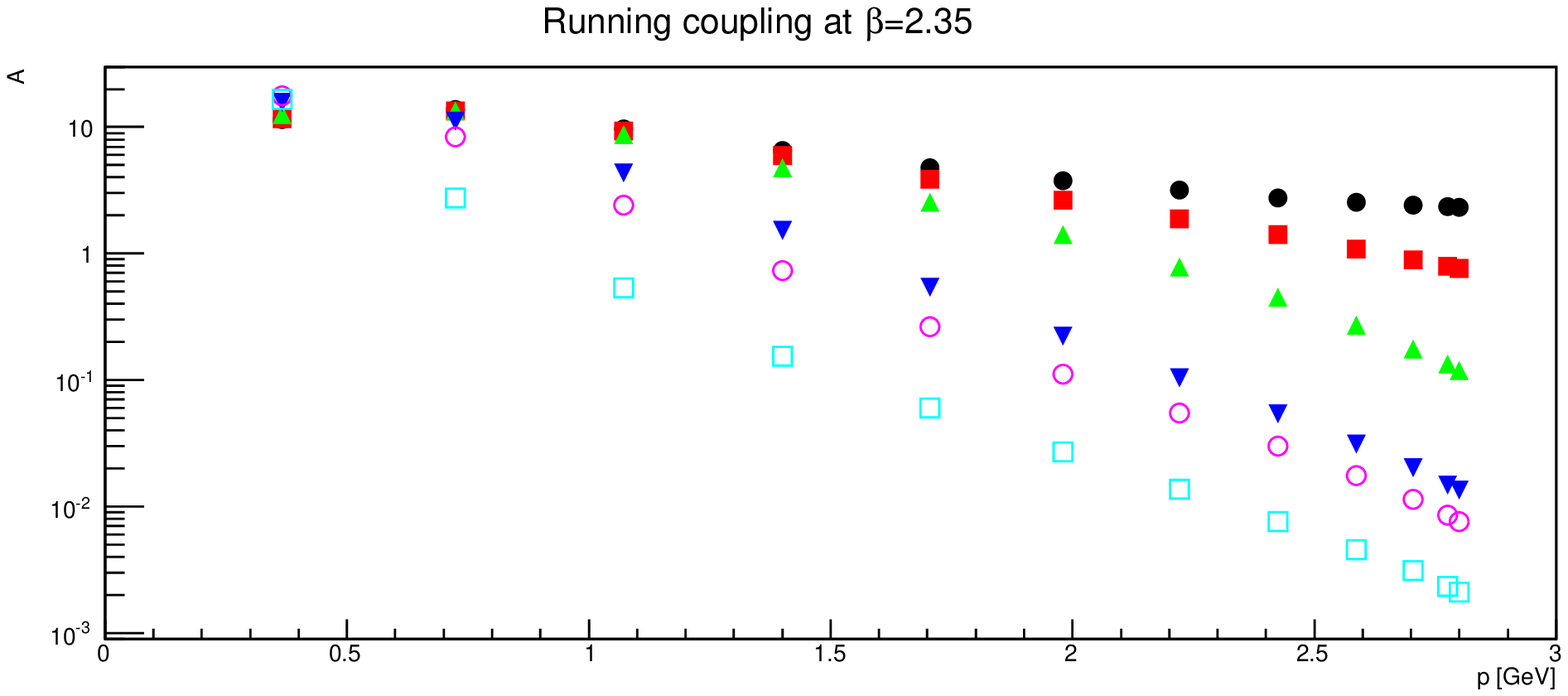}\\
\includegraphics[width=\linewidth]{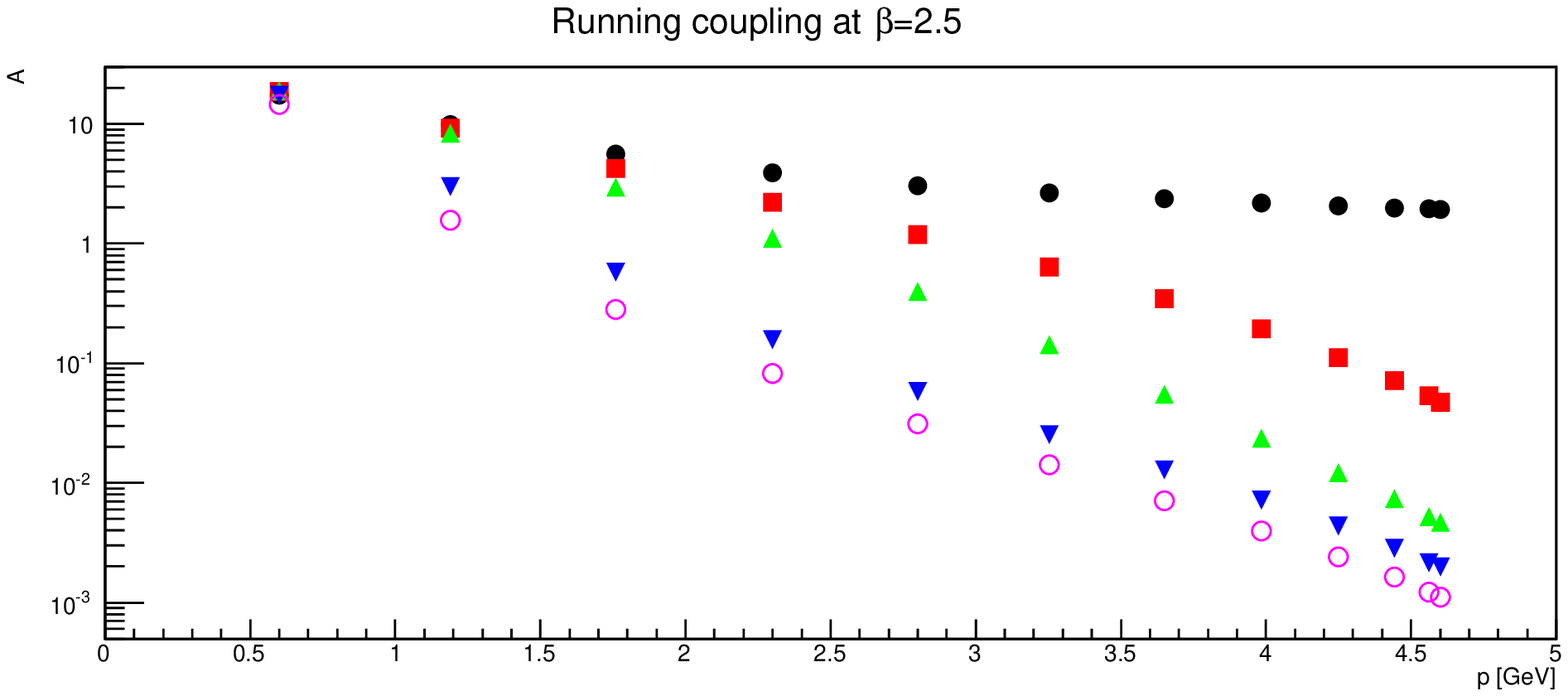}
\caption{\label{fig:alpha}The momentum-dependent part $A$ of the running coupling \pref{alpha} for different numbers of APE sweeps. The top panel is for a discretization of $a=0.21$ fm, the middle panel for $a=0.14$ fm, and the bottom panel for $a=0.087$ fm. All results from $24^4$ lattices.}
\end{figure}

The running coupling is a useful quantity to assess how the structure of the fields affects the interaction strength. In Landau gauge, a running coupling can be defined in the miniMOM scheme by \cite{vonSmekal:2009ae,Alkofer:2000wg}
\be
\alpha(p)=\alpha(\mu) p^6 D_G(p,\mu)^2 D(p,\mu)=\alpha(\mu)A(p,\mu)\label{alpha},
\ee
\no where $\alpha(\mu)$ normalizes the coupling, making $\alpha(p)$ independent of $\mu$. Since this is an irrelevant (scheme-dependent) overall factor, here only the dimensionless product $A=p^6 D_G^2 D$ will be investigated, as the main interest is the change of momentum-dependence under smearing. This dependence on smearing is shown in figure \ref{fig:alpha}. The results agree with what could have been inferred from figures \ref{fig:gp2}-\ref{fig:gp5} and \ref{fig:ghp2}-\ref{fig:ghp5}: At small momenta, the decrease in the ghost propagator cannot compensate the enhancement of the gluon propagator. Thus, the running coupling becomes a steeply increasing function at mid momentum, before eventually bending down again, if the volume is sufficiently large to reach small enough momenta. At the same time, the ultraviolet suppression of the gluon propagator dominates the effect at large momenta, since the ghost propagator is almost not altered. As a consequence, the ultraviolet interactions are strongly suppressed, reflecting itself in a ultraviolet suppressed running coupling.

\subsection{Residual configuration results}\label{sresi}

In principle, it is possible to view the smeared links $U_\mu^s$ just as part of the original links $U_\mu$, with the remainder part defined as
\be
U_\mu^r=U_\mu U_\mu^{s\dagger}\label{resconf}.
\ee
\no This implies that for an unsmeared configuration $U_\mu^r$ is the unit matrix, and the corresponding gluon propagator vanishes therefore identical and the ghost propagator is the one of a free particle.

Since matrix multiplication is not commutative, this is not a unique decomposition. Nonetheless, similar decompositions have been used in the past \cite{Greensite:2004ur,Gattnar:2004bf,Langfeld:2002dd,Boucaud:2003xi,Langfeld:2001cz,Quandt:2010yq} to characterize the contents of the part of the configuration removed under smearing and/or other operations to isolate the topological content. Based on the assumption that smearing removes only the ultraviolet fluctuations, the corresponding propagators have usually been found to be approximately the perturbative or tree-level ones.

\begin{figure}
\includegraphics[width=0.5\linewidth]{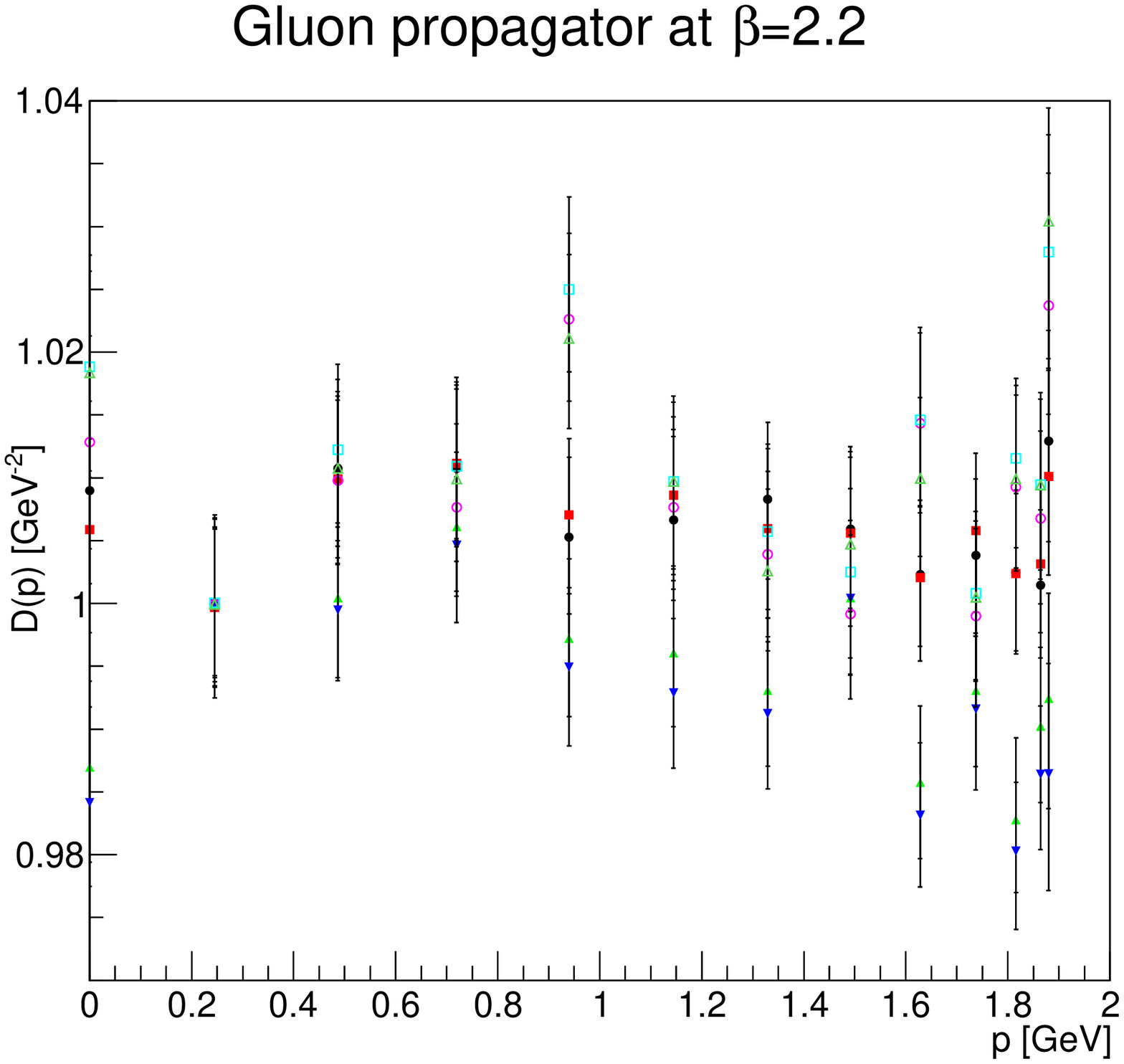}\includegraphics[width=0.5\linewidth]{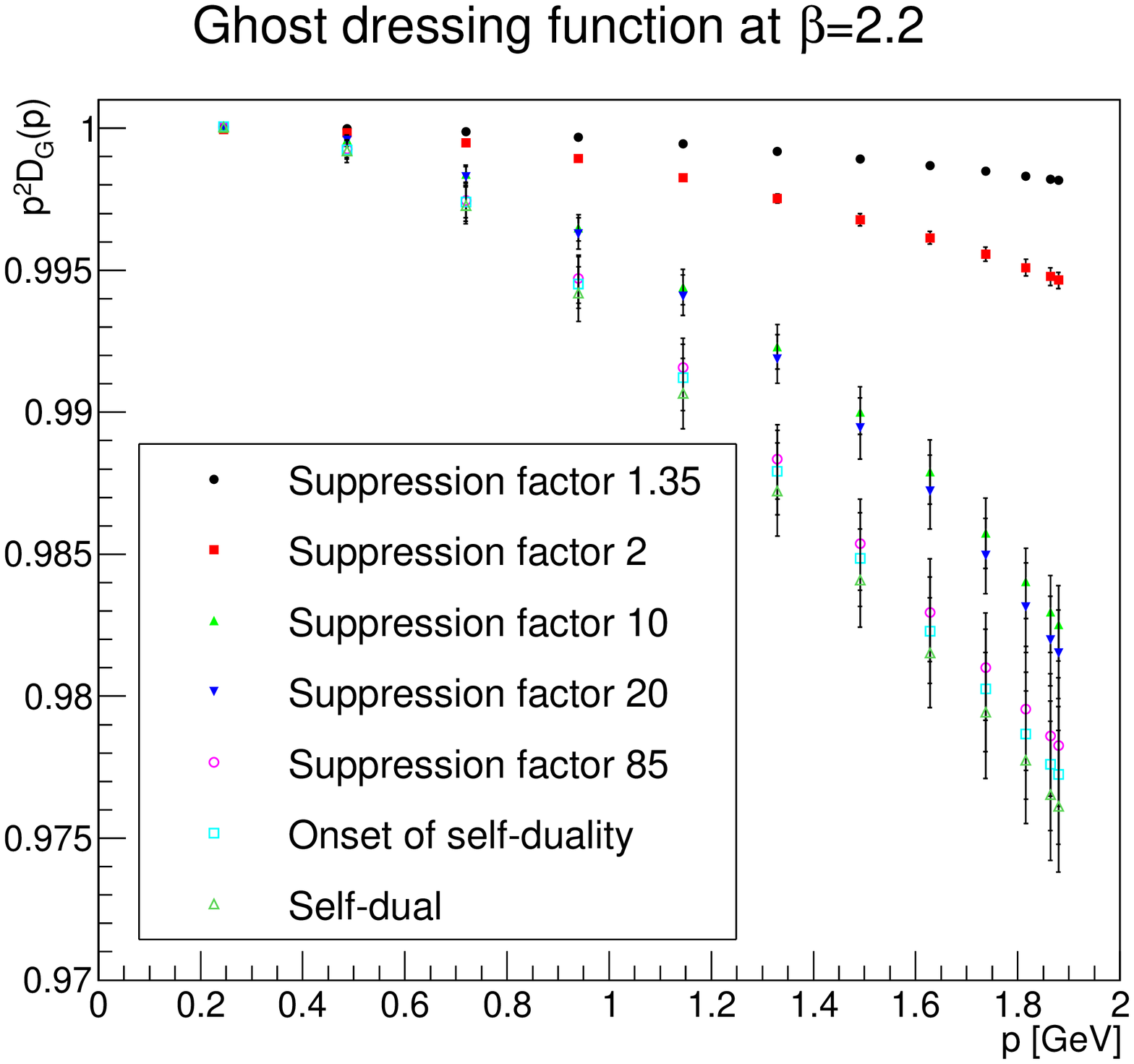}\\
\includegraphics[width=0.5\linewidth]{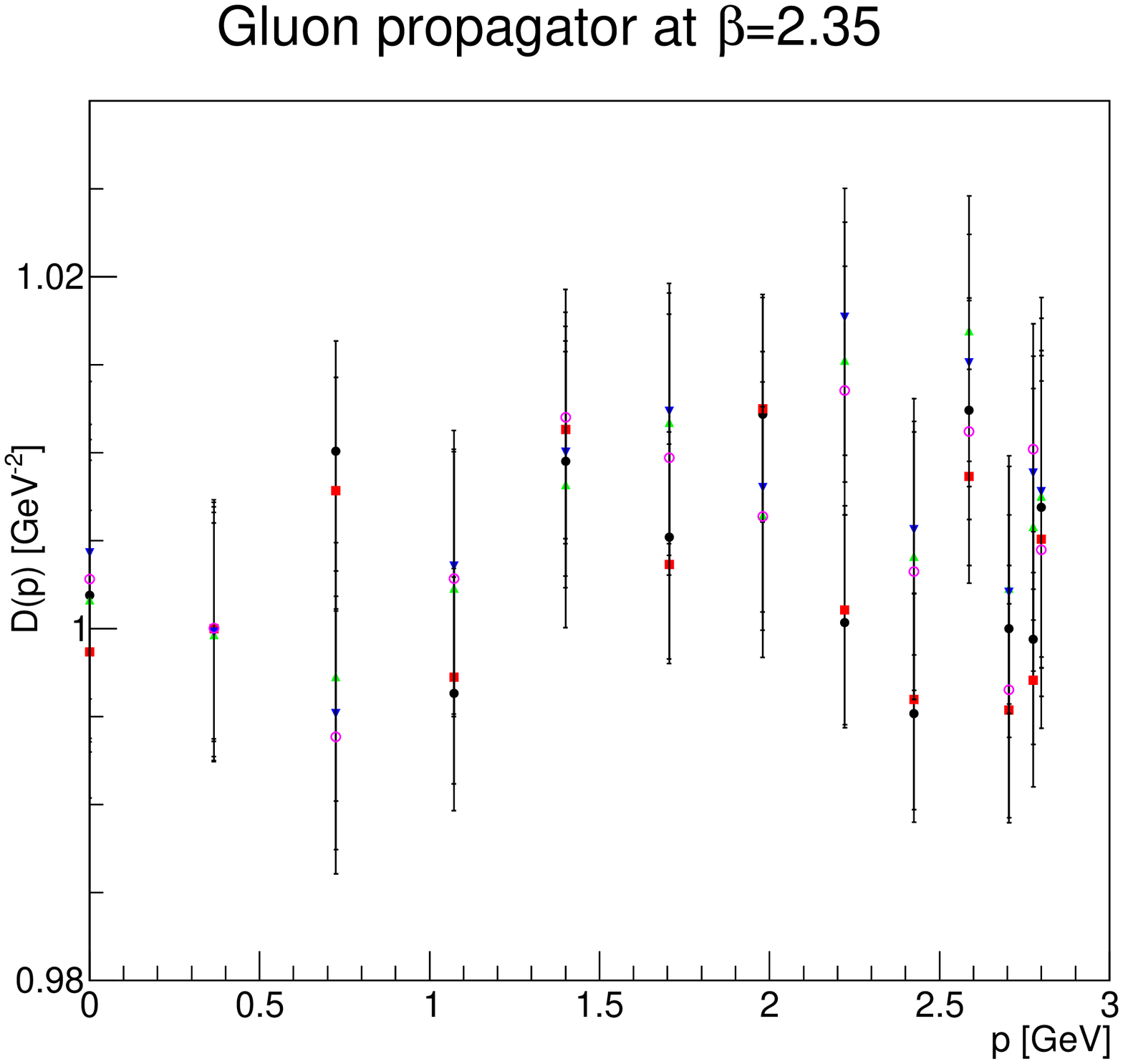}\includegraphics[width=0.5\linewidth]{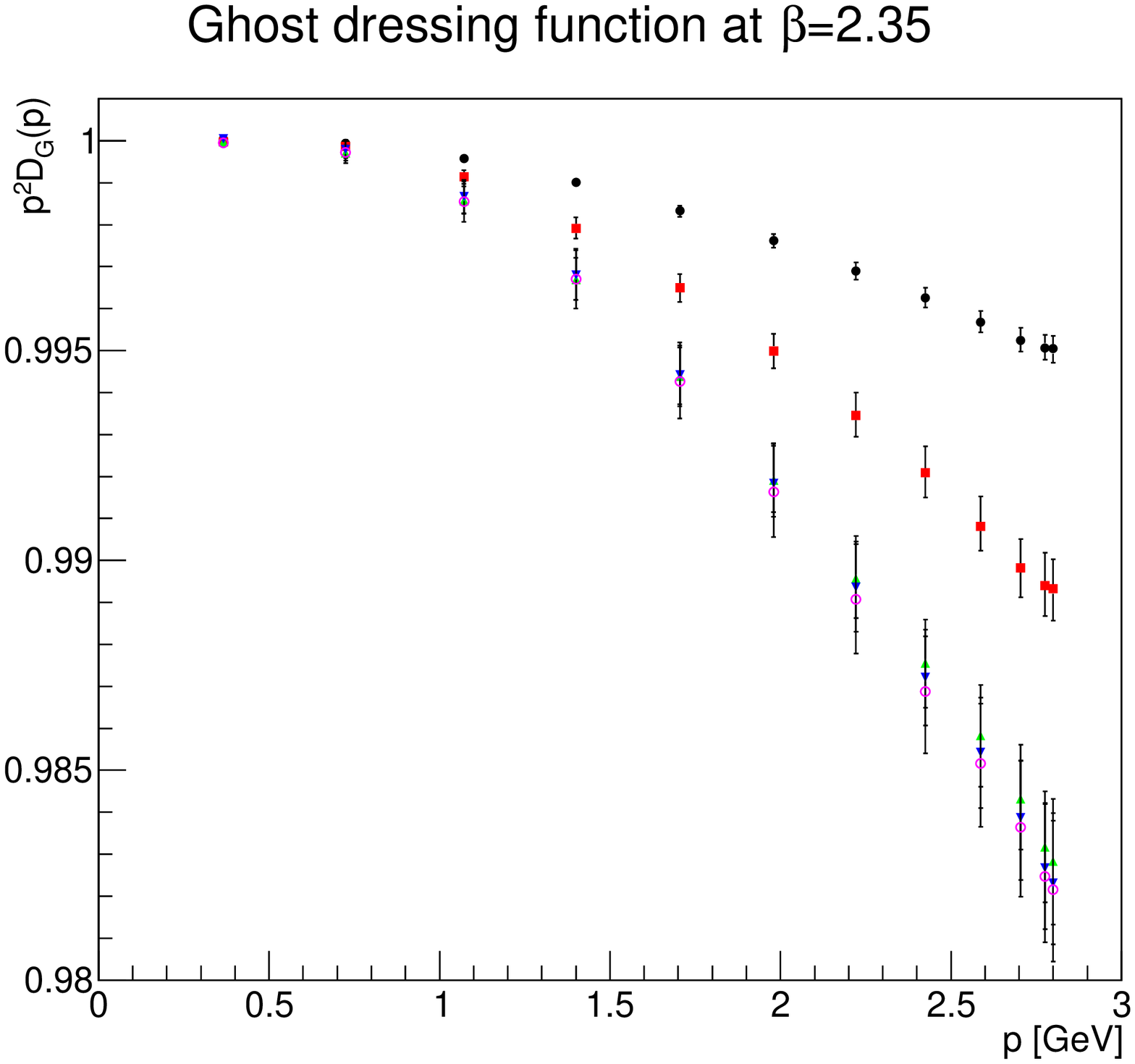}\\
\includegraphics[width=0.5\linewidth]{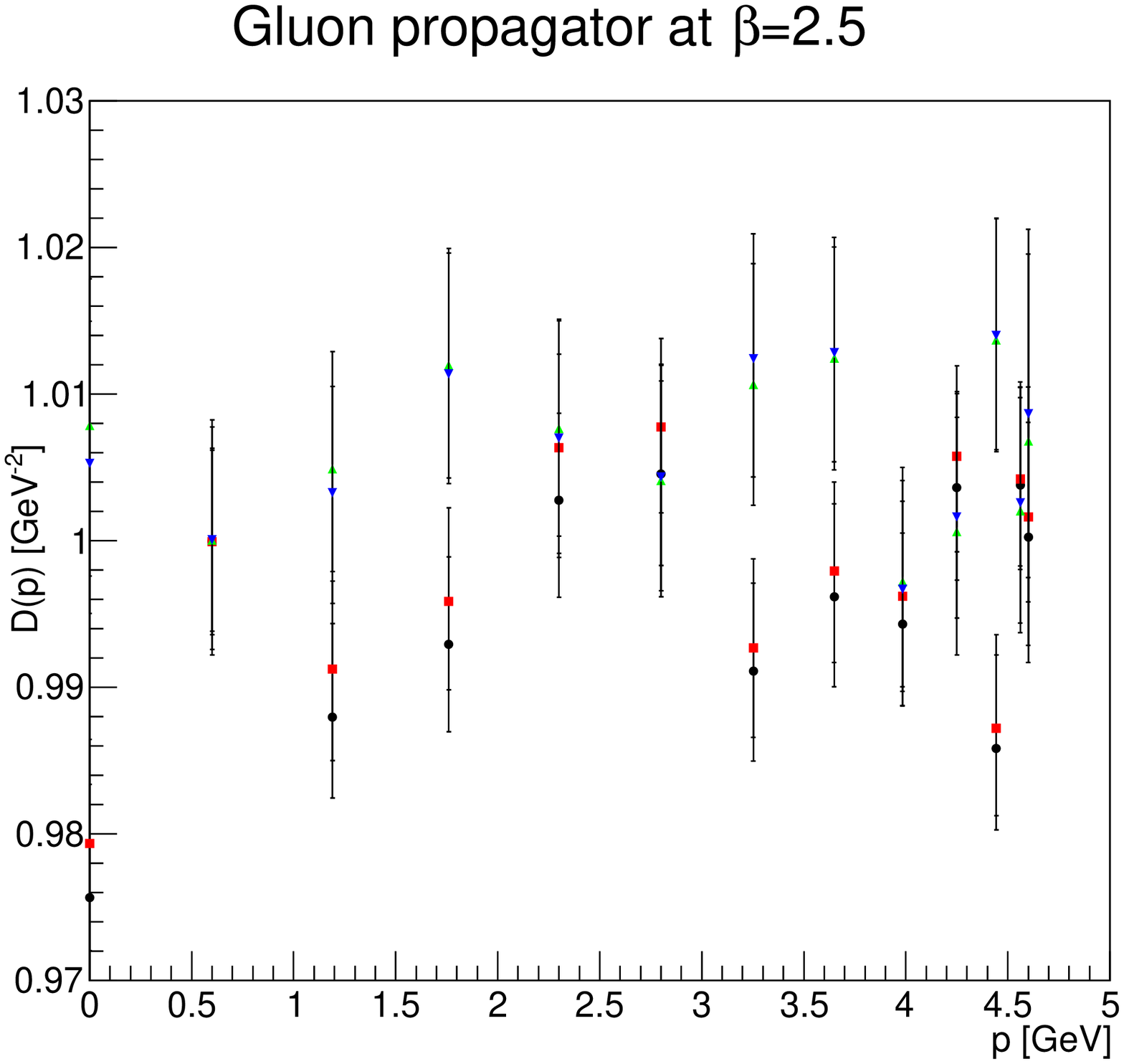}\includegraphics[width=0.5\linewidth]{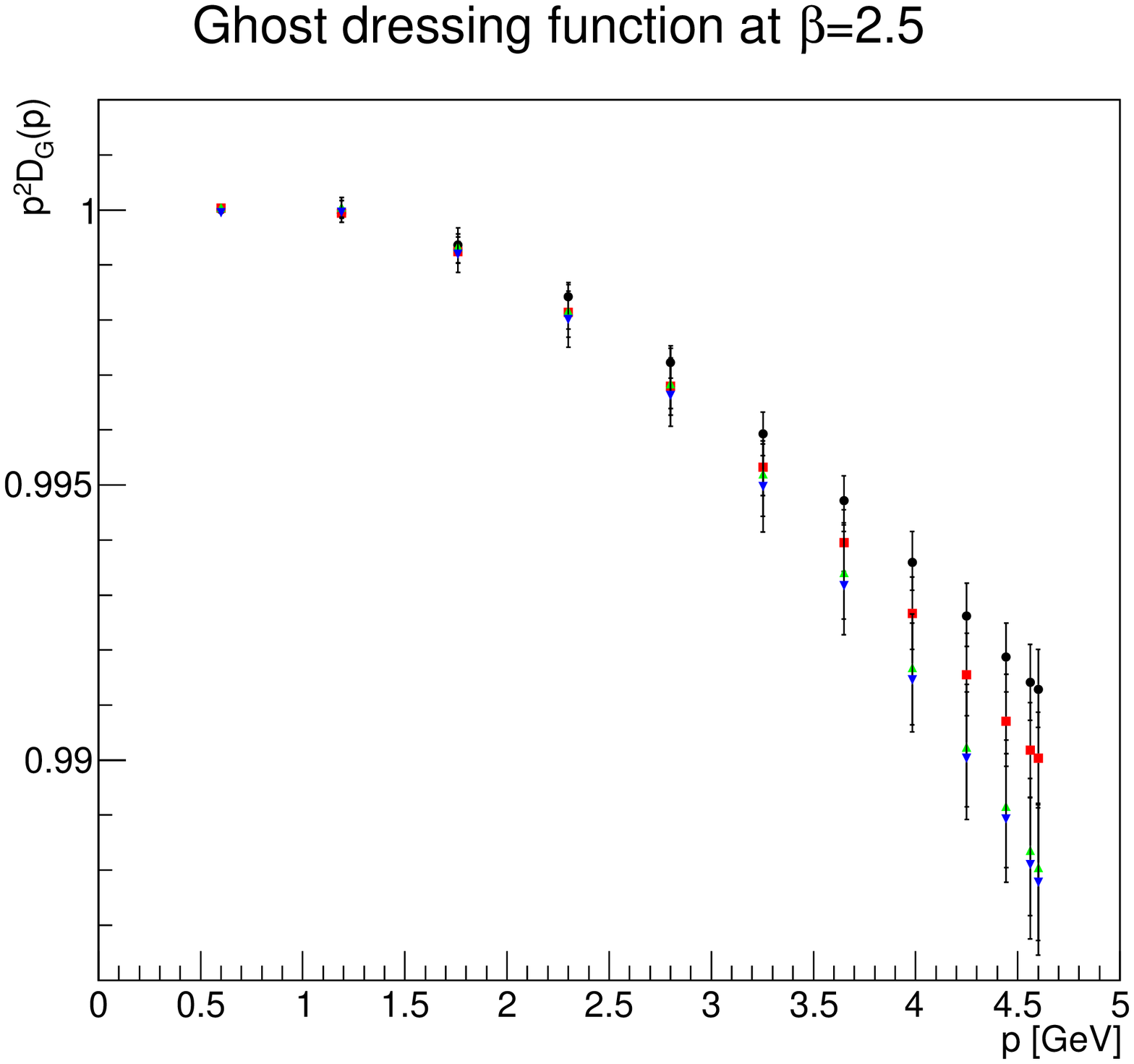}
\caption{\label{fig:res}The gluon propagator (left panels) and ghost dressing function (right panels) of the residual configurations \pref{resconf} for different numbers of APE sweeps. The gluon propagator and the ghost dressing function have been normalized to one at the lowest non-vanishing momentum. The top panels are for a discretization of $a=0.21$ fm, the middle panels for $a=0.14$ fm, and the bottom panels for $a=0.087$ fm. All results from $24^4$ lattices. Note the scales.}
\end{figure}

In figure \ref{fig:res} the corresponding propagators and dressing functions for different numbers of APE sweeps are shown. Surprisingly, these function agree within a few percent, after appropriate normalization and after a certain amount of smearing. The overall normalization is a decreasing function of the number of APE sweeps, which appears to tend to a finite value for larger and larger numbers of APE sweeps, though this has not been studied in detail. Also, there is a significant impact, especially for the ghost dressing function, of the discretization. The finer and smaller the lattice, the closer the ghost dressing function at the same amount of suppression moves towards the tree-level behavior.

The results are as expected. The gluon propagator is essentially momentum independent, and thus the one of a random variable, $\langle A_\mu^a(x) A_\nu^a(y)\rangle=\delta^{ab}\delta_\mn\delta(x-y)$. Consequently, the dressing function increases quickly with momentum. Such a behavior is also seen in the Schwinger function: It is zero, within error, except for $x=y$. The ghost dressing function shows only a mild deviation from the one of a free particle at large momentum for an increasing level of smearing. Otherwise, it is essentially that of a free, massless particle.

\section{A speculative interpretation}\label{sspec}

The results shown here are for a very limited amount of different volumes and discretizations, due to the computational costs involved. Given the sensitivity of the investigated correlation functions to both types of lattice artifacts \cite{Maas:2011se}, any interpretation should only be made with the requirement that it has to be confirmed for larger volumes, and thus remains currently speculative. It should also be kept in mind that any other way of selecting Gribov copies than the minimal Landau gauge used here may yield different results.

Assuming for a moment that there will be no qualitative change for larger volumes leads then to the following interpretation: Also for propagators smearing leads to a suppression of ultraviolet physics. Consequently, the gluon propagator is ultraviolet suppressed. The corresponding residual propagator is the one of a free random variable. The ghost propagator shows at large momenta the one of a free field, and thus decouples from the gluonic fluctuations. At the same time, the infrared and mid-momentum behavior is non-trivial. Thus, the long-distance behavior of the propagators appears to be dominated by self-dual (topological) configurations, in line with the arguments in \cite{Boucaud:2003xi,Maas:2005qt,Maas:2006ss}. This also confirms the idea that topological scenarios and the physics captured by the infrared behavior of correlation functions are just two facets of the same underlying physics \cite{Alkofer:2006fu}.

A somewhat surprising result is the appearance of a range of momenta at the hadronic scales where the gluon propagator appears to behave like $1/p^4$. This would be a welcomed effect, as such a contribution makes it much more easier to understand where the gluon propagator encodes information like the Wilson string. Taking this for granted, this would imply that by smearing the physical irrelevant, gauge-fixing-dominated infrared part is shifted to small momenta, and the relevant mid-momentum physics, which is thus generated by the topological excitations, is made evident. This would imply, in a very simplified manner, that the unsmeared gluon propagator consists out of three parts: A ultraviolet part, a strongly-interacting $1/p^4$ part, and an infrared finite part due to the gauge-fixing process. These are schematically multiplicatively superimposed like
\be
D(p)=\frac{p^4}{p^2+\gamma^2}\times\frac{1}{p^4}\times D_p(\gamma^2+p^2)\nn,
\ee
\no where $\gamma$ is the Gribov parameter \cite{Vandersickel:2012tg}, and $D_p$ is just a perturbative part. Smearing removes the ultraviolet part, and reduces the Gribov parameter, and thus emphasizes the $1/p^4$ behavior\footnote{This interpretation is reminiscent of the stochastic vacuum picture \cite{Dosch:1988ha,Nachtmann:1996kt}. I am grateful to Reinhard Alkofer for pointing this out.}. An interesting challenge for this interpretation is, what would happen in two dimensions under smearing. Because of the absence of dynamics, the $1/p^4$ part is absent, and the gluon propagator vanishes hence in the infrared. If this is true, the gluon propagator in two dimensions under smearing would not develop an enhancement. However, due to geometric confinement, already present in two-dimensional QED, this effect may be obscured.

Besides this conceptual insight, this has also practical implications for non-lattice calculations of correlations functions, especially for functional methods. Any approximation and/or truncation made in such calculations can in principle remove part or all of the physics due to topological configurations. The investigations here now provide a necessary condition for the correct inclusion of this infrared physics. Only if the infrared and the mid-momentum behavior of the correlation functions is correct, the contributions from topology is included. This is not sufficient, at least if only a finite number of correlation functions is determined, since it appears possible that the infrared behavior could also be reproduced without the correct physics.

On the other hand, there is a remarkable implication in return. It has been explicitly tested that for many physical observables that the far infrared behavior of correlation function appears to be irrelevant \cite{Blank:2010pa,Luecker:2009bs,Fischer:2009gk,Fischer:2009jm,Costa:2010pp}. Since this part appears to be retained under smearing, this implies that this unphysical information is still encode in the topological fluctuations as well, besides any physical information. Given the limited amount of results here, also this statement has to be taken as being speculative.

\section{Conclusions}\label{sconc}

Summarizing, it is found that for the limited set of lattice settings studied here the low-momentum behavior of correlation functions is qualitatively conserved under APE smearing. That implies that likely the infrared part of correlation functions carry indeed the imprint of topological configurations with them, emphasizing that both aspects are only two facets of the same underlying physics. Furthermore, no pronounced dependence of the correlation functions on (net-)topological charge is found. This implies that the correlation functions are not strongly affected if lattice simulation algorithms get stuck in a sector of fixed topological charge.

Of course, a precise quantitative investigation will require much larger lattices, especially in the asymptotic domain \cite{Maas:2011se}. However, such an investigation will require substantially more computational resources, and therefore will only be possible when these become available in the future. Still, the present results provide besides the physical insights the necessary requirement that analytical determinations of correlation functions need to be correct at small and intermediate momenta to include topological effects. On the other hand, since the configurations are purely self-dual/topological after a sufficient number of APE sweeps, this implies that correlation functions can capture the physics of such configurations, and no artificial additional introduction of them into, e.\ g., functional methods is required.\\

\no{\bf Acknowledgments}

I am grateful to {\v S}tefan Olejn\'ik for many helpful discussions in the early stages of this work, and to the referee for very helpful and constructive suggestions. This project was supported by the DFG under grant numbers MA 3935/1-2 and MA 3935/5-1 and the FWF under grant numbers P20330 and M1099-N16. Simulations were performed on the HPC cluster at the Universities of Jena and Graz. I am grateful to the corresponding HPC teams for the very good performance of both clusters. The ROOT framework \cite{Brun:1997pa} has been used in this project.

\bibliographystyle{bibstyle}
\bibliography{bib}


\end{document}